\documentclass[
twocolumn,
floatfix,
aps,
prd,
showpacs,
amsmath, 
amssymb, 
nofootinbib,
]{revtex4}

\usepackage{graphicx}
\usepackage{bm}
\usepackage{epsf}

\newcommand{\beq}{\begin{equation}}
\newcommand{\eeq}{\end{equation}}
\newcommand{\beqn}{\begin{eqnarray}}
\newcommand{\eeqn}{\end{eqnarray}}
\newcommand{\pa}{\partial}

\begin{document}

\title{Bar-mode instability of rapidly spinning black hole 
in higher dimensions: \\
Numerical simulation in general relativity}

\author{Masaru Shibata and Hirotaka Yoshino}

\affiliation{Yukawa Institute for Theoretical Physics, 
Kyoto University, Kyoto, 606-8502, Japan}

\affiliation{Department of Physics, University of Alberta, 
Edmonton, Alberta, Canada T6G 2G7}

\begin{abstract}
Numerical-relativity simulation is performed for rapidly spinning
black holes (BHs) in a higher-dimensional spacetime of special
symmetries for the dimensionality $6 \leq d \leq 8$.  We find that
higher-dimensional BHs, spinning rapidly enough, are dynamically
unstable against nonaxisymmetric bar-mode deformation and
spontaneously emit gravitational waves, irrespective of $d$ as in the
case $d=5$ \cite{SY09}.  The critical values of a nondimensional spin
parameter for the onset of the instability are $q:=a/\mu^{1/(d-3)}
\approx 0.74$ for $d=6$, $\approx 0.73$ for $d=7$, and $\approx 0.77$
for $d=8$ where $\mu$ and $a$ are mass and spin parameters. Black
holes with a spin smaller than these critical values ($q_{\rm crit}$)
appear to be dynamically stable for any perturbation.  Longterm
simulations for the unstable BHs are also performed for $d=6$ and
7. We find that they spin down as a result of gravitational-wave
emission and subsequently settle to a stable stationary BH of a spin
smaller than $q_{\rm crit}$. For more rapidly spinning unstable BHs,
the timescale, for which the new state is reached, is shorter and
fraction of the spin-down is larger. Our findings imply that a highly
rapidly spinning BH with $q > q_{\rm crit}$ cannot be a stationary
product in the particle accelerators, even if it would be formed as a
consequence of a TeV-gravity hypothesis. Its implications for the
phenomenology of a mini BH are discussed.
\end{abstract}
\pacs{04.25.D-, 04.30.-w, 04.40.Dg}

\maketitle

\section{Introduction}

Clarifying formation and evolution processes of mini black hole (BH)
in higher-dimensional spacetimes has become an important issue since a
possibility of BH formation in huge particle accelerators was pointed
out. If our space is a 3-brane in a higher-dimensional flat spacetime
of spacetime dimensionality $d \ge 6$ \cite{ADD98} or in an anti-de
Sitter (AdS) spacetime of $d \ge 5$ \cite{RS99}, the Planck energy
could be of $O({\rm TeV})$ that may be accessible with particle
accelerators in operation, the CERN Large Hadron Collider (LHC). In
the presence of the extra dimensions, BHs of very small mass energy
$\agt$~TeV may be produced during the particle collision in the
accelerators because the true Planck energy may be as low as TeV
scale.

A hypothetical phenomenology of a BH produced in the huge particle
accelerator was first discussed in \cite{BHUA,GT02} (see
\cite{reviews} for reviews).  According to this standard scenario, a
mini BH evolves in the following manner: During the high-energy
particle collision of a sufficiently small impact parameter and of
energy sufficiently higher than the Planck energy,
two particles will merge to form a deformed BH, and then, it settles
to a quasistationary state after emission of gravitational
waves. The typically assumed timescale for gravitational-wave emission
is $\sim 10r_+/c$ (about 10 times of the dynamical timescale) where
$r_+$ and $c$ are the horizon radius and speed of light,
respectively. The quasistationary BH will be subsequently evaporated
by the Hawking radiation \cite{Hawking}, because of
quantum-field-effects in a curved spacetime.  Much effort has been
devoted to calculating the greybody factor in the Hawking radiation
for improving the prediction of signals in the particle-collision
experiments \cite{greybody,Ida1,Ida2,Ida3,greybody-rotating,KKKPZ09}
(see also \cite{related} for related issues).

By contrast, the analyses for BH formation after the particle
collision and subsequent evolution by gravitational radiation reaction
in higher-dimensional spacetime have not been done yet [but see
  \cite{HEADON,SOY,SCPBHY,CP} for studies in the four-dimensional (4D)
  case].  These phases are expected to be described well in the
context of general relativity \cite{GR04}, but due to its highly
nonlinear nature, any approximation breaks down. Obviously,
numerical-relativity simulation is the unique approach for studying
this phase.

One of the important issues has been to clarify what type of BH is
formed and whether it is stable or not. In the 4D case, 
any stationary BH formed
in vacuum has to be a Kerr BH (neglecting the electric charge of the
BH) because of the uniqueness theorem (e.g., \cite{unique} for
review), and the Kerr BH has been proven to be stable
\cite{Kerr-stability,leaver,whiting} (but see \cite{FN} for remaining
issues for a perfect proof). These facts 
strongly constrain the possible scenario for mini BH formation and its
subsequent evolution.  By contrast, there is no uniqueness theorem and
no proof for the stability of higher-dimensional BHs (but see \cite{IM04} for
uniqueness of 5D BHs of the spherical horizon
topology).  As a result, the standard scenario described above (i.e.,
formation, evolution by subsequent gravitational-wave emission, and
evaporation by the Hawking radiation) is quite uncertain.

A mini BH, if it is formed as a result particle collision in
a higher-dimensional spacetime, will have only one spin parameter
associated with the orbital motion.  This can restrict the possibility
for the type of the formed BH. However, even in this case, one cannot
restrict the BH type in higher dimensions. For example, in the 5D
case, there are many types of BHs: e.g., usual {\em Kerr-type} BH
(Myers-Perry BH \cite{MP86}) for which surface of the event horizon is
${\rm S}^3$, fat and thin {\em black rings} for which surface has a
ring-like shape \cite{ER03}. Recently, black di-ring \cite{RING} and
black-saturn \cite{SATURN} solutions were derived as well (see also
\cite{Other-solutions} for other exact solutions of 5D black objects
but with two spin parameters).  Several authors \cite{Ida1,YN04,YS07}
discuss a possibility of the black-ring formation in particle
collisions.  Although it is still an open question, the analysis of
apparent horizons in \cite{YN04} indicates that black-ring formation
is not very likely in two point-particle system.

Higher-dimensional BHs with a high spin parameter are known to be
unstable against {\em axisymmetric} perturbations.  Emparan and Myers
\cite{EM03} speculated that rapidly spinning BHs with the spacetime
dimensionality $d \geq 6$ are subject to the Gregory-Laflamme
instability \cite{GL93}, because they have a high degree of
ellipticity (i.e., the black membrane limit).  Very recently, Dias
{\it et al.} and Murata {\it et al.} indeed showed, by a linear
perturbation analysis, that rapidly spinning BHs for $6 \le d\le 9$
are unstable against axisymmetric multiple-ring-like deformation
\cite{Dias,Murata} (see also related papers \cite{Kerr-related}).

On the other hand, little is known for the stability of spinning BHs
against {\it nonaxisymmetric} perturbations. Emparan and
Myers~\cite{EM03} also speculated, based on a thermodynamic argument,
that rapidly spinning BHs with a sufficiently high spin may be
unstable against nonaxisymmetric perturbation for $d \ge 5$, because
the horizon area (the so-called entropy) of a rapidly spinning BH is
often smaller than that of two boosted Schwarzschild BHs with the same
total energy and angular momentum. However, the correspondence between
the thermodynamical and dynamical instabilities has not been well
established.  Authors in \cite{hydrodynamics} guess, based on a
hydrodynamic/gravity correspondence argument, that a nonaxisymmetric
instability may occur for spinning BHs with spin smaller than the
critical value for the onset of axisymmetric instabilities.  However,
a rigorous and quantitative analysis is absent in these studies.  To
strictly clarify what the criterion for the onset of nonaxisymmetric
instabilities is and how the instabilities occur and proceed, we have
to solve Einstein's equation.  Recently, we performed a
numerical-relativity simulation for 5D spinning BHs for the first
time, and found that BHs spinning rapidly are dynamically unstable
against nonaxisymmetric bar-mode deformation if the spin parameter
satisfies the condition $q:=a/\mu^{1/2} \agt 0.87$ (see Sec.~II for
the definitions of $q$, $a$, and $\mu$) \cite{SY09}: An interesting
fact is that the critical value we found is close to the value
predicted by Emparan and Myers~\cite{EM03}, $\approx 0.85$, suggesting
that their argument relying on BH thermodynamics may be reliable
(although we show in this paper that this is not the case for $d \geq
6$).

In this paper, we study the stability of rapidly spinning BHs with
dimensionality $d \geq 6$ using a new numerical-relativity code
{\small SACRA-ND}, which is extended from {\small SACRA} \cite{SACRA}
and {\small SACRA5D} \cite{SY09}. In this code, Einstein's equation
for higher-dimensional spacetimes of any dimensionality for $d \geq 5$
is fully solved without imposing axial symmetry.  Thus, in this work,
we do not have to assume that the amplitude of the perturbation from
the background axisymmetric BH solution is small nor restrict
attention to an axisymmetric perturbation.  The merits of this
approach are that (i) a wide variety of instabilities can be
investigated with no ambiguity and with no approximation (except for
finite-difference approximation); (ii) the final fate after the onset
of the instabilities can be determined because the amplitude of the
perturbation from the background BH solution does not have to be
small; and (iii) the criterion for the onset of the instabilities is
{\em quantitatively} determined.  In this paper, we focus on the six-,
seven-, and eight-dimensional (6D, 7D, and 8D) BHs of single spin
parameter because such a class of BH is a possible outcome in the
particle accelerators. We shall show, for the first time, that the
rapidly spinning BHs are dynamically unstable against bar-mode
deformation irrespective of the dimensionality. We, furthermore,
evolve the 6D and 7D unstable BHs for a long time until they settle to
a new stable state. This enables to quantitatively clarify a possible
(classical) evolution process of an unstable mini BH for the first
time.

The paper is organized as follows. In Sec.~II, the formulation for
solving Einstein's evolution equation in higher dimensions is
described. In Sec.~III, a method for evolving Einstein's equation in
the spacetime of a special symmetry is summarized. In Sec.~IV, we
present the numerical results focusing on the criterion for the onset
of the bar-mode dynamical instability of rapidly spinning BHs and the fate
after the onset of the instability. We will show that the rapidly spinning
BHs, which have a spin larger than the critical value, are unstable
against spontaneous emission of quadrupole gravitational waves. We
will also show that all the unstable BHs, considered in this paper,
evolve as a result of gravitational radiation reaction, settling to a
stable BH of spin smaller than the critical value. In Sec.~V, the
conditions for the spontaneous gravitational-wave emission and
associated spin-down process are clarified. Section VI is devoted to a
summary and discussion. In particular, implications of our results for
the phenomenology of a mini BH are discussed in detail.

In Secs.~I -- VI~B, we adopt the units in which $c=1$.  Only in
Sec.~VI~C, the natural units $c=\hbar=1$, where $\hbar$ is the Planck
constant, are adopted.  $d$ denotes the spacetime dimension with
$n:=d-4$ being the number of the extra dimension, and  $G_d$ denotes the
$d$-dimensional gravitational constant.  The Cartesian coordinates
$(x, y, z, w_q)$ are used for the space coordinates: $x^a=(x,y,z)$
denote the usual three dimensional coordinates, and $w_q~(q=1$ -- $n)$
the coordinates of extra dimension. $t$ denotes the time.  Indices
$i$, $j$, $k$, and $l$ denote the general spatial coordinates.

\section{Formulation}

We consider a vacuum higher-dimensional spacetime of
$\mathrm{SO}(n+1)$ symmetry for which the line element is written as
\beqn
&&ds^2=-(\alpha^2 - \beta_k \beta^k)dt^2 + 2 \beta_k dx^k dt \nonumber \\
&& \hskip 2cm  
+\gamma_{ab}dx^a dx^b + \gamma_{nn} d\Omega^2_n, \label{eq1}
\eeqn
where $\alpha$ is the lapse function, $\beta^k$ the shift vector, 
$\gamma_{ij}$ the space metric, and $d\Omega^2_n$ the 
line element of $n$-dimensional unit sphere. 
$\gamma_{nn}$ is a conformal factor for the extra-dimensional 
metric components. $x^a$ denotes $(x, y, \rho)$ where $\rho$ is a 
radial coordinate 
\beqn
\rho=\sqrt{z^2+\sum_{q=1}^n w_q^2}~~. 
\eeqn
The geometric quantities in this symmetric spacetime depend only on
$t$, $x$, $y$, and $\rho$.  (Note that by $\mathrm{SO}(n+1)$ symmetry,
we imply that the subspace of $(z, w_1, \cdot, w_n)$ coordinate
directions is isotropic.)

The line element of a Myers-Perry BH of single spin (i.e., the
Kerr-type BH) in the Boyer-Lindquist-type coordinates is \cite{MP86}
\beqn
&&ds^2=-dt^2+{\mu \over \hat r^{d-5} \Sigma}(dt-a\sin^2\theta d\varphi)^2 
+{\Sigma \over \Delta} d\hat r^2 + \Sigma d\theta^2
\nonumber \\
&& ~~~~~~~+(\hat r^2 + a^2) \sin^2\theta d\varphi^2
+\hat r^2 \cos^2\theta d\Omega_n^2, \label{kerr}
\eeqn
where $\mu$ and $a$ are mass and spin parameters, respectively, 
$\Sigma:=\hat r^2 +a^2\cos^2\theta$, and $\Delta:=\hat r^2 +a^2 -
\mu \hat r^{5-d}$. Note that the mass and angular momentum of this BH are
\beqn
&&M={(d-2)\Omega_{d-2} \mu \over 16\pi G_d},
\label{MP-parameter-mu}
\\
&&J={2 \over d-2} M a,
\label{MP-parameter-a} 
\eeqn
where $\Omega_{d-2}$ is the area of $(d-2)$-dimensional unit sphere,
i.e., $\Omega_{d-2}=2\pi^{(d-1)/2}/\Gamma[(d-1)/2]$.  Thus, the line
element denoted by Eq.~(\ref{eq1}) includes the spinning BH solution
(\ref{kerr}), and nonstationary, nonaxisymmetric deformed states can
be described as well.

In numerical simulation, we adopt the Cartesian coordinates $(x, y, z,
w_q)$ instead of the curvilinear coordinates; e.g., for $d=7$, the
relations between $(z, w_1, w_2, w_3)$ and radial and angular coordinates
$(\rho, \psi, \varphi_1, \varphi_2)$ are
\beqn
z  &=&\rho \cos \psi, \\ 
w_1&=&\rho \sin \psi \cos \varphi_1, \\ 
w_2&=&\rho \sin \psi \sin \varphi_1 \cos \varphi_2, \\ 
w_3&=&\rho \sin \psi \sin \varphi_1 \sin \varphi_2.
\eeqn
Then, Einstein's evolution equation is solved in the Cartesian 
coordinates using the so-called cartoon method \cite{cartoon,YS09}. 
Namely, we solve the equations in the $(x, y, z)$ hyperplane; 
the hyperplane of $\psi=\varphi_i=0$. The method used in the present 
work will be described in the next section. 

To solve Einstein's evolution equation in the Cartesian coordinates,
we adopt a multidimensional version of the 
Baumgarte-Shapiro-Shibata-Nakamura (BSSN) formalism
\cite{BSSN,YS09} (see also \cite{Zilhao2010}
for a different formalism).  We rewrite the line element in the form,
\beqn
ds^2=-(\alpha^2 - \beta_k \beta^k)dt^2 + 2 \beta_k dx^k dt + 
\chi^{-1}\tilde \gamma_{ij}dx^i dx^j, \label{BSSN}
\eeqn
where $\chi=[{\rm det}(\gamma_{ij})]^{-1/(d-1)}$ is a conformal factor and 
the conformal spatial metric $\tilde \gamma_{ij}$ satisfies the 
condition ${\rm det}(\tilde \gamma_{ij})=1$. In addition, 
we define the following quantities from the extrinsic curvature $K_{ij}$, 
\beqn
&& \tilde A_{ij} := \chi \left( K_{ij}-{1 \over d-1}\gamma_{ij}K \right),\\
&& K := K_{ij} \gamma^{ij},
\eeqn
as well as an auxiliary variable 
\beq
\tilde \Gamma^i := - \pa_j \tilde \gamma^{ij}.
\eeq 
Then, the variables $(\chi, \tilde \gamma_{ij}, K, \tilde A_{ij}, 
\tilde \Gamma^i)$ are evolved solving the following equations \cite{YS09}:
\beqn
&&(\partial_t-\beta^k\partial_k)\chi
=\frac{2}{d-1}\chi\left(\alpha K-\partial_i\beta^i\right), 
\label{evolution-chi} \\
&& (\partial_t
-\beta^k\partial_k)\tilde{\gamma}_{ij}=
-2\alpha\tilde{A}_{ij}
+\tilde{\gamma}_{ik}\partial_j\beta^k
+\tilde{\gamma}_{jk}\partial_i\beta^k \nonumber \\
&& \hskip 4cm -\frac{2}{d-1}\tilde{\gamma}_{ij} \partial_k\beta^k, 
\label{evolution-gammaij}\\
&&(\partial_t -\beta^k\partial_k)K = -D_i D^i \alpha
+\alpha\left(\tilde{A}^{ij}\tilde{A}_{ij}+\frac{K^2}{d-1}\right),~~~
\label{evolution-K} \\
&&(\partial_t-\beta^k\partial_k)\tilde{A}_{ij}
=\chi\left[-(D_i D_j\alpha)^{\rm TF} +\alpha R_{ij}^{\rm TF}\right]
\nonumber \\
&& ~~~~~~~ +\alpha\left(K\tilde{A}_{ij}-2\tilde{A}_{ik}\tilde{A}^k_j\right)
\nonumber \\
&& ~~~~~~~ +\tilde{A}_{ik}\partial_j\beta^k
+\tilde{A}_{kj}\partial_i\beta^k
-\frac{2}{d-1} \tilde{A}_{ij} \partial_k\beta^k,
\label{evolution-Aij} \\
&&
(\partial_t-\beta^k\partial_k)\tilde{\Gamma}^i
=-2\tilde{A}^{ij}\partial_j\alpha
-\tilde{\Gamma}^j\partial_j\beta^i
\nonumber \\
&& ~
+\frac{2}{d-1}
\tilde{\Gamma}^i \partial_j\beta^j
+\frac{d-3}{d-1}\tilde{\gamma}^{ik}\pa_k \pa_j \beta^j
+\tilde{\gamma}^{jk} \pa_j \pa_k \beta^i
\nonumber \\
&& ~
+2\alpha\biggl[
\tilde{\Gamma}^{i}_{jk}\tilde{A}^{jk} 
-\frac{d-2}{d-1}\tilde{\gamma}^{ij}\pa_j K
-\frac{(d-1)}{2}\frac{\pa_j \chi}{\chi}\tilde{A}^{ij}
\biggr]. \label{evolution-Gammai}
\eeqn
Here, $D_i$ and $R_{ij}$ denote the covariant derivative 
and the Ricci tensor with respect to $\gamma_{ij}$, 
and TF denotes taking the trace-free part. 

For evaluating $R^{\rm TF}_{ij}$, we first decompose $R_{ij}$ as
\beqn
R_{ij}=\tilde R_{ij} + R^\chi_{ij},
\eeqn
where $\tilde R_{ij}$ is the Ricci tensor with respect to 
$\tilde \gamma_{ij}$ and we write it in the following form
\beqn
&&\tilde{R}_{ij}=
-\frac12\tilde{\gamma}^{kl}\tilde{\gamma}_{ij,kl}
+\frac12\left(\tilde{\gamma}_{ki}\partial_j\tilde{\Gamma}^k+
\tilde{\gamma}_{kj}\partial_i\tilde{\Gamma}^k\right)
\nonumber \\
&& ~~~~~~~ 
-\frac12\biggl[
 (\pa_l \tilde \gamma_{ik}) \pa_j \tilde \gamma^{kl}
+(\pa_l \tilde \gamma_{jk}) \pa_i \tilde \gamma^{kl}
-\tilde \Gamma^l \pa_l \tilde \gamma_{ij} \biggr]
\nonumber \\
&& ~~~~~~~ 
-\tilde \Gamma^l_{ik} \tilde \Gamma^k_{jl}.
\label{bar-Ricci}
\eeqn
Here, $\tilde \Gamma^i_{jk}$ is the Christoffel symbol with respect to 
$\tilde \gamma_{ij}$. 
$R^{\chi}_{ij}$ is sum of the terms associated with $\chi$, 
\beqn
&& R_{ij}^{\chi}=
\frac{(d-3)}{2\chi} \tilde D_i \tilde D_j \chi
-\frac{(d-3)}{4}\frac{(\tilde D_i \chi) \tilde D_j \chi}{\chi^2}
\nonumber \\
&& ~~~~~~~
+\tilde{\gamma}_{ij} \biggl[{1 \over 2\chi} \tilde D_k \tilde D^k \chi
-{d-1 \over 4 \chi^2}(\tilde D_k \chi) \tilde D^k \chi \biggr].~~~
\label{Ricci-chi}
\eeqn
Then, we obtain  
\beqn
&&R_{ij}^{\rm TF}=\tilde R_{ij}-{1 \over d-1} \tilde \gamma_{ij} 
\tilde\gamma^{kl} \tilde R_{kl} \nonumber \\
&& ~~~~~ + {d-3 \over 2 \chi} 
\left[\tilde D_i \tilde D_j \chi - {1 \over d-1} \tilde \gamma_{ij}
\tilde D_k \tilde D^k \chi\right] \nonumber \\
&& ~~~~~ - {d-3 \over 4\chi^2} \left[
(\tilde D_i \chi) \tilde D_j \chi - {1 \over d-1} \tilde \gamma_{ij}
(\tilde D_k \chi) \tilde D^k \chi \right]. ~~~
\eeqn

For the 4D case, the so-called puncture gauge conditions are known to
be robust for evolving BH spacetime with the BSSN formalism
\cite{BHBH}. This is also the case for the higher-dimensional
spacetime \cite{YS09}. However, the freely chosen coefficients in this
gauge have to be carefully determined for the stable and longterm
evolution. Specifically, we choose the equations in the form
\beqn
&&\pa_t \alpha= -1.5 \alpha K, \label{gauge1}\\
&&\pa_t \beta^i = 0.3B^i,\label{gauge2}\\
&&\pa_t B^i=(\pa_t-\beta^k \pa_k) \tilde \Gamma^i - \eta_B \mu^{-1/(d-3)} B^i,
\label{gauge3}
\eeqn
or 
\beqn
\pa_t B^i=\pa_t \tilde \Gamma^i - \eta_B \mu^{-1/(d-3)} B^i,
\label{gauge3a}
\eeqn
where $B^i$ is an auxiliary function, and $\eta_B$ is a nondimensional
constant for which we give different values for different number of
$d$ because a small value of $\eta_B$ is not allowed for a high number
of $d$ or for a high spin for achieving the longterm stable numerical
evolution.  For $d=5$, we gave $\eta_B=1$ in the previous work
\cite{SY09}.  In the present work, we employ $\eta_B=2$ -- 5 for
$d=6$, 3 -- 5 for $d=7$, and 8 for $d=8$.  For unstable BHs, a large
value of $\eta_B$ is favored in performing a longterm simulation until
the growth of the instability saturates and subsequently the
deformation damps. We tried to use both Eqs.~(\ref{gauge3}) and
(\ref{gauge3a}), and found that both of them work well as far as BH
spin is not extremely large. For a very high spin ($q \agt 1$),
however, Eq.~(\ref{gauge3a}) works better than Eq.~(\ref{gauge3}) for
a longterm stable simulation.

\section{Modified cartoon method}

We solve Einstein's evolution equation in the spacetime of
$\mathrm{SO}(n+1)$ symmetry with $n=d-4$ using the Cartesian
coordinates $(x,y,z,w_1,\cdots,w_n)$ with the BSSN formalism. The
$\mathrm{SO}(n+1)$ symmetry is imposed for the $(z,w_1,\cdots,w_n)$
subspace (i.e., the subspace of ($z, w_1,..., w_n$) coordinate
directions is assumed to be isotropic).  The method is qualitatively
the same as those described for 5D spacetime \cite{YS09}.  Namely, we
solve the equations in the $(x, y, z)$ hyperplane (i.e.,
$w_1=w_2=\cdots=w_n=0$ hyperplane), and derivatives with respect to
these coordinates are evaluated by a straightforward finite
differencing. On the other hand, the derivatives with respect to the
extra-dimensional coordinates, $w_q~(q=1 \cdots n)$, are evaluated
using symmetry relations. In the previous paper \cite{YS09}, we
adopted the original prescription in the cartoon method often employed
for axisymmetric spacetimes \cite{cartoon}: We prepare 4 additional
grid points for the fifth coordinate, $w$, i.e., $2\pm \Delta w$ and
$\pm \Delta w$. Then, the values for all the geometric quantities at
these grid points are determined using the symmetry relation
associated with the Killing vector $\pa/\pa \psi$ where
$\psi=\tan^{-1}(w/z)$, and then, the derivatives with respect to $w$
are evaluated using a fourth-order finite differencing.

This original method is quite simple for code implementation.  However,
it is memory-consuming in higher-dimensional simulations, because we
have to increase the grid number by a factor of five whenever the
number of extra dimension is increased. To save the memory used in
computation, we employ a different method in which the derivatives
with respect to $w_q$ are replaced to those with respect to $z$
without preparing additional grid points for the extra-dimensional
direction. In the following, we describe the new method denoting the
scalar, vector, and tensor by $Q$, $Q^i$, and $Q_{ij}$, respectively, 
and decomposing the subscripts $i,j$ into $a$, $b=x$, $y$, $z$ and 
$w_q$ ($q=1$ -- $n$). 

In the $(x, y, z)$ hyperplane, the following relations hold because of
the assumed isotropy:
\begin{equation}
Q_{ww}:= Q_{w_1w_1} = \cdots = Q_{w_nw_n}
\label{diagonal-components-relation}
\end{equation}  
and
\begin{equation}
Q^{w_q}=Q_{aw_q} = Q_{w_qw_r} = 0 ~~(q\neq r).
\label{offdiagonal-components-relation}
\end{equation}
The derivatives with respect to $x^a$
of Eq.~(\ref{offdiagonal-components-relation})
are trivially zero. For other derivatives, 
the following relations hold: For the scalar quantities,
\beqn
&& Q_{,w_q}=0, ~~ Q_{,w_q w_r}=(Q_{,z}/z)\delta_{qr}, 
\label{cart1a}
\eeqn
for the vector quantities,
\beqn
&& Q^a_{~,w_q}=Q^a_{~,b w_q}=Q^{w_p}_{~~,w_q w_r}=0, 
\nonumber\\
&& Q^{w_q}_{~~,w_r}=\left(Q^z/z\right)\delta_{qr}, 
~~ Q^{w_q}_{~~,w_r a}=\left(Q^z/z\right)_{,a}\delta_{qr}, 
\nonumber\\
&& Q^A_{~,w_q w_r}=\left(Q^A_{~,z}/z\right)\delta_{qr}, 
\nonumber\\
&& Q^z_{~,w_q w_r}=\left(Q^z/z\right)_{,z}\delta_{qr}, 
\label{cart1b}
\eeqn
and for the tensor quantities, 
\beqn
&& Q_{ab,w_q}=Q_{w_p w_q,w_r} = 0,~~ 
Q_{A w_q,w_r}=(Q_{Az}/z)\delta_{qr}, \nonumber\\
&& Q_{zw_q, w_r}=[(Q_{zz}-Q_{ww})/z]\delta_{qr},\nonumber \\
&& Q_{AB,w_q w_q} = Q_{AB,z}/z, 
~~ Q_{Az,w_q w_q} = (Q_{Az}/z)_{,z}, \nonumber \\ 
&& Q_{zz,w_q w_q} = Q_{zz,z}/z - (2/z^2)(Q_{zz}-Q_{ww}), \nonumber \\ 
&& Q_{w_q w_q,w_q w_q}=Q_{ww,z}/z
+ ({2}/{z^2})(Q_{zz}-Q_{ww}), \nonumber \\ 
&& Q_{w_q w_q,w_r w_r}=Q_{ww,z}/z ~~(q\neq r). 
\label{cart1c}
\eeqn 
Here, $A, B=x$ or $y$ (not $z$) and $\delta_{qr}$ is the Kronecker's
delta. We did not evaluate $Q_{ij,aw_q}$ and $Q_{ij,w_qw_r}$ $(q\neq
r)$ even if they are not vanishing, because they do not 
appear in the BSSN equations with $\mathrm{SO}(n+1)$ symmetry: 
Note that the second derivatives appear only in the term
$-(1/2)\tilde{\gamma}^{kl} \pa_k \pa_l \tilde{\gamma}_{ij}$ of
Eq.~(\ref{bar-Ricci}) and $\tilde{\gamma}^{kl}$ satisfies the 
relation of Eq.~(\ref{offdiagonal-components-relation}).  We also
do not have to evaluate $Q_{aw_q,ij}$ and $Q_{w_qw_r, ij}$ $(q\neq r)$, 
because they appear only in the $a w_q$ and $w_q w_r$ components of
Eq.~(\ref{evolution-Aij}) with Eq.~(\ref{bar-Ricci}) which do not have 
to be evolved, as mentioned later. 

Some of the above prescriptions can be used only for $z \not=0$ 
in computer because of the presence of the terms associated with
$1/z$, although these terms are actually regular. Thus for $z=0$, 
the following relations, which are found from $\mathrm{SO}(n+1)$ symmetry, 
are employed: 
\beqn
&& Q_{,w_q w_q}=Q_{,zz}, \nonumber \\
&& Q^{w_q}_{~~,w_q}=Q^z_{~,z}, ~~~Q^A_{~,w_q w_q}=Q^A_{~,zz},
~~~Q^z_{~,w_q w_q}=0,\nonumber \\
&& Q_{AB,w_q w_q}=Q_{AB,zz}, ~~~ Q_{Az,w_q w_q}=0, \nonumber \\ 
&& Q_{A w_q,w_q}=Q_{Az,z}, ~~~ Q_{zw_q, w_q}=0,\nonumber \\
&& Q_{zz,w_q w_q}=Q_{ww,zz},~~~Q_{w_q w_q,w_q w_q}=Q_{zz,zz}, \nonumber \\ 
&& Q_{w_q w_q,w_r w_r}=Q_{w_q w_q,zz}. \label{cart2}
\eeqn

The final remark is on the treatment of the advection terms such as 
$\beta^k\pa_k \tilde \gamma_{ij}$: Because $\beta^{w_q}=0$ for the
$w_q=0$ hyperplane in the present case, the advection terms associated
with $\beta^{w_q}$ is always vanishing in the computational domain 
chosen in our method. 

With these prescriptions, all the derivatives associated with the
extra-dimensional coordinates can be replaced to the
finite-differencing terms with respect to $x^a$ or with no finite
differencing. It is worthy to note that total amount of computational
operation is only slightly larger than that for the 3+1 case.

Because of the relations (\ref{diagonal-components-relation}) and
(\ref{offdiagonal-components-relation}) that follow from
$\mathrm{SO}(n+1)$ symmetry, the implementation for the
higher-dimensional contribution can be even simplified: We have only
to evolve the scalar equations, the ``$a$'' components of the vector
equations, and the ``$ab$'' components and (one of) the $w_q w_q$
components of the tensor equations (i.e., the equations for
$\tilde{\gamma}_{ij}$ and $\tilde{A}_{ij}$) of the BSSN formalism.
Therefore, we need to increase only one component for $\tilde
\gamma_{ij}$ and $\tilde A_{ij}$ irrespective of the dimensionality.
In the evolution equation, we often have terms such as $\beta^i_{~,i}$
or $\tilde \gamma^{kl}\tilde \gamma_{ak,l}$, which are evaluated by
\beqn
&& \beta^i_{~,i}=\beta^a_{~,a} + n \beta^{w_1}_{~,w_1},\\
&& \tilde \gamma^{kl}\tilde \gamma_{ak,l}=
\tilde \gamma^{cd}\tilde \gamma_{ac,d} 
+ n\tilde \gamma^{w_1w_1}\tilde \gamma_{a w_1,w_1},
\eeqn
where the prescriptions shown in Eqs.~(\ref{cart1a}) -- (\ref{cart2})
are used for evaluating the second terms in the right-hand side.
These facts imply that once a 5D code is implemented, it is quite
straightforward to extend it to a code for $d \geq 6$, even when we do
not employ a curvilinear coordinate system~\cite{Zilhao2010}. 

\section{Numerical simulation}

\subsection{Setting and Methodology}

We prepare 6D -- 8D Myers-Perry BHs of single spin parameter (see
Eq.~(\ref{kerr}) for the line element in the Boyer-Lindquist
coordinates) as the initial condition.  Because the Boyer-Lindquist
coordinates are not suitable for the Cauchy evolution of BH
spacetimes, we transform the radial coordinate introducing a
quasiradial coordinate $r$; we rewrite the line element in the form
(see \cite{KP} for the 4D case)
\beqn
&&ds^2=-dt^2+{\mu \over \hat r^{d-5}\Sigma}(dt-a\sin^2\theta d\varphi)^2 
\nonumber \\
&& ~~~~~~~~
+\Phi (dr^2 + r^2 d\theta^2)
\nonumber \\
&& ~~~~~~~~
+(\hat r^2 + a^2) \sin^2\theta d\varphi^2
+\hat r^2 \cos^2\theta d\Omega_n^2,~~ \label{kerr2}
\eeqn
where $\Phi$ is a conformal factor for the $(r, \theta)$ plane. 
Namely, the two-dimensional metric for $(r, \theta)$ is written in 
a conformally flat form, and the following relations are satisfied: 
\beqn
&& \Phi^{1/2} dr = \pm (\Sigma/\Delta)^{1/2} d\hat r,\\
&& \Phi^{1/2} r = \Sigma^{1/2}.
\eeqn
Then, $r$ is defined by 
\beqn
r = r_h \exp\left[\pm \int_{r_+}^{\hat r} {dR \over \sqrt{R^2 +a^2-\mu R^{5-d}}}
\right], \label{quasi}
\eeqn 
where the plus and minus signs are adopted for the regions $r \geq
r_h$ and $r\leq r_h$, respectively.  $r_+$ is the horizon radius in
the Boyer-Lindquist coordinates and a positive root of $r_+^2+a^2=\mu
r_+^{5-d}$. $r_h$ is the horizon radius in the quasiradial coordinate, 
which is determined by the condition $r=\hat r$ for $\hat r
\rightarrow \infty$.  A fourth-order numerical integration is
performed for the integral of Eq.~(\ref{quasi}) using the Bode's rule 
\cite{recipe} because it cannot be analytically integrated in general.

With this transformation, the $t=$const hypersurface becomes spacelike
everywhere for $ 0 \leq r < \infty$, and furthermore, the singularity,
which is originally located at $\hat r=0$, is excluded. More
specifically, the spacelike hypersurface has an inversion symmetry
with respect to $r=r_h$ hypersphere (i.e., this hypersphere is the
worm-hole throat), and thus the ``point'', $r=0$, represents the
spacelike infinity of (say) another world beyond the horizon.
Although this point is not a physical singularity, it becomes a
coordinate singularity, because $\Phi$ is proportional to $r^{-4}$
irrespective of the number of $d$. However, the puncture approach
[specifically, appropriate choice of the conformal factor in numerical
simulation, $\chi=[{\rm det}(\gamma_{ij})]^{-1/(d-1)}$, and choice of
the puncture gauge (\ref{gauge1}) -- (\ref{gauge3})] enables to stably
evolve a BH spacetime with no difficulty \cite{BHBH,YS09}.

The condition for the onset of instabilities for spinning BHs depends
on a nondimensional spin parameter. Because the stability does not
depend on the magnitude of $\mu$, all dimensional quantities can be
scaled out of the problem appropriately normalizing them by using
$\mu$: Specifically, $\mu^{1/(d-3)}$ has the dimension of length and
time in the $c=1$ units (and is often referred to as $r_s$,
e.g.,\cite{Ida3}), and thus, we should define a nondimensional spin as
\beqn
q:= {a \over \mu^{1/(d-3)}}.
\eeqn
The stability of a Myers-Perry BH depends only on this quantity. 
Note that this is different from $a_*:=q/r_+$ which is often 
used as another nondimensional spin parameter \cite{GT02,Ida3}. 

Numerical simulation is performed in the $(x, y, z)$ coordinates, and 
thus, the initial condition is prepared by performing a 
coordinate transformation from $(r, \theta, \varphi)$ to $(x, y, z)$. 
In addition, we add a small bar-mode perturbation to the 
conformal factor $\chi$ as
\beqn
\chi=\chi_0 \left[
1 + A{x^2 - y^2 \over \mu^{2/(d-3)}} \exp\left(-{r^2 \over 2r_h^2}\right) 
\right],
\eeqn 
where $\chi_0$ is the nonperturbed solution and $A$ the initial
perturbation amplitude, $A \ll 1$. We focus here on the bar mode
because it is often the most relevant unstable mode for
self-gravitating, dynamically unstable rotating systems such as a
rotating star \cite{CH69}.  We chose the value of $A$ for a wide range
from $10^{-6}$ to $0.02$ and found that for unstable BHs, the growth
rate of the unstable mode does not depend on the initial magnitude of
$A$. Taking into account this result, in a longterm evolution of
unstable BHs for studying the nonlinear growth of the unstable mode,
we choose a relatively large value of $A=0.005$ or 0.02 to save
computational costs. It should be also pointed out that the numerical
error accumulates with time and a significant resolution is needed to
suppress it.  This implies that if a simulation was started with a
very small value of $A$, quite expensive computational costs would be
required to accurately follow the nonlinear growth of the
perturbation.

Initially, the lapse function is chosen as $\alpha=\chi$. With this
modification, $\alpha$ becomes positive except for $r=0$ where
$\alpha=0$ (near $r=0$, $\alpha$ is proportional to $r^4$ with this
choice).  On the other hand, the shift vector is not modified; we
choose the same shift as in the Myers-Perry BH initially.

We solve Einstein's evolution equation in the BSSN formalism using a
new code, {\small SACRA-ND}, in which an adaptive mesh refinement
(AMR) algorithm is implemented in the same manner as in {\small SACRA}
\cite{SACRA}.  Because we solve the higher-dimensional equations only
with the $(x, y, z)$ coordinates, the methods of interpolation,
extrapolation, and evolution for the grid structure in {\small
  SACRA-ND} are totally the same as those in {\small SACRA}: All the
spatial derivatives with respect to $x^a=(x, y, z)$ are evaluated
using a centered fourth-order finite differencing except for the
advection term such as $\beta^k \pa_k \chi$ for which a fourth-order
upwind scheme is adopted.  The time evolution is carried out using the
standard fourth-order Runge-Kutta method \cite{recipe}. The reader may
refer to \cite{SACRA} for details about the numerical methods.

For the AMR scheme, we prepare 6 refinement levels in the present
numerical simulation. As a test, we performed several simulations
enlarging the computational domain with 7 refinement levels while the
grid resolution in the finest level is unchanged. We found that the
results shown in Sec.~IV are essentially independent of the location
of the outer boundary. Computational domain for each refinement level
has a half cubic shape which covers $[-L_l:L_l]$ for $x$ and $y$, and
$[0:L_l]$ for $z$; we assume the equatorial plane symmetry.  Here,
$L_l$ denotes the location of the refinement boundary for the $l$-th
refinement level with $l=0$ -- $N_l$ and $N_l=5$ in the present
work. We note that in {\small SACRA} and {\small SACRA-ND}, six grid
points are prepared outside the refinement boundaries for buffer zone
in each refinement level, following \cite{BHBH2}.  The relation
$L_l=2L_{l+1}$ is imposed for {\small SACRA-ND}.  Irrespective of $l$,
we assign $2N+1$ vertex-centered grid points for $[-L_l:L_l]$ where
$N$ was chosen to be 30, 40, and 50 for checking convergence: We
monitored the violation of the Hamiltonian and momentum constraints,
conservation of gravitational mass, and conservation of the area and
spin of BHs (for stable model), and checked that the convergence with
improving the grid resolution is achieved.  The grid spacing is $L_l/N$
and $L_{N_l}$ is chosen to be $1.2 \mu^{1/(d-3)}$ irrespective of the
number of $d$. For the best resolved run, the apparent horizon radius
is covered by 30 -- 40 grid points for rapidly spinning BHs: We note
that the initial coordinate radii of the BH horizon,
$r_h/\mu^{1/(d-3)}$, are in the range 0.584 -- 0.630 ($q=1.143$ -- 0,
minimum at $q \sim 1$) for $d=6$, 0.750 -- 0.707 ($q=1.013$ -- 0) for
$d=7$, and 0.812 -- 0.758 ($q=0.832$ -- 0) for $d=8$, respectively.
Because of our choice of the puncture gauge, this coordinate radius
increases by a factor of 1.3 -- 1.5 during evolution.

A steep gradient always appears for geometric quantities near the
origin for rapidly spinning BHs in the puncture gauge, and this often
causes a problem for stably evolving the BHs, although such region is
not important for studying the stability of the BH at all.  Thus, we
employ a very simple excision method for discarding this
region. Specifically, for $r \leq r_{\rm ex}$, we set $\tilde
\gamma_{ij}=\delta_{ij}$ and $\tilde A_{ij}=K=0$. Other quantities,
$\alpha$, $\beta^k$, $\chi$, and $\tilde \Gamma^i$, are solved without
any prescription. In this work, $r_{\rm ex}$ is chosen to be
$0.3\mu^{1/(d-3)}$ or $0.4\mu^{1/(d-3)}$ which is typically $\sim 30$
-- 40\% of the apparent horizon radius; thus, the excised region is
well inside the BH horizon. Because no information escapes from the BH
horizon and there are many grid points covering the inside of apparent
horizon, the results are insensitive to the choice of $r_{\rm ex}$ as
far as it is sufficiently small. We note that this method is
acceptable for the case that an unstable BH deforms by a moderate
degree (for $\eta \alt 0.5$; see Sec.~\ref{sec:nr} for definition of
$\eta$). For a highly nonlinear deformation, however, we will have to
develop other prescriptions for handling the steep gradient near the
central region, which is beyond the scope of this paper.

\begin{figure*}[t]
\epsfxsize=3.2in
\leavevmode
(a)\epsffile{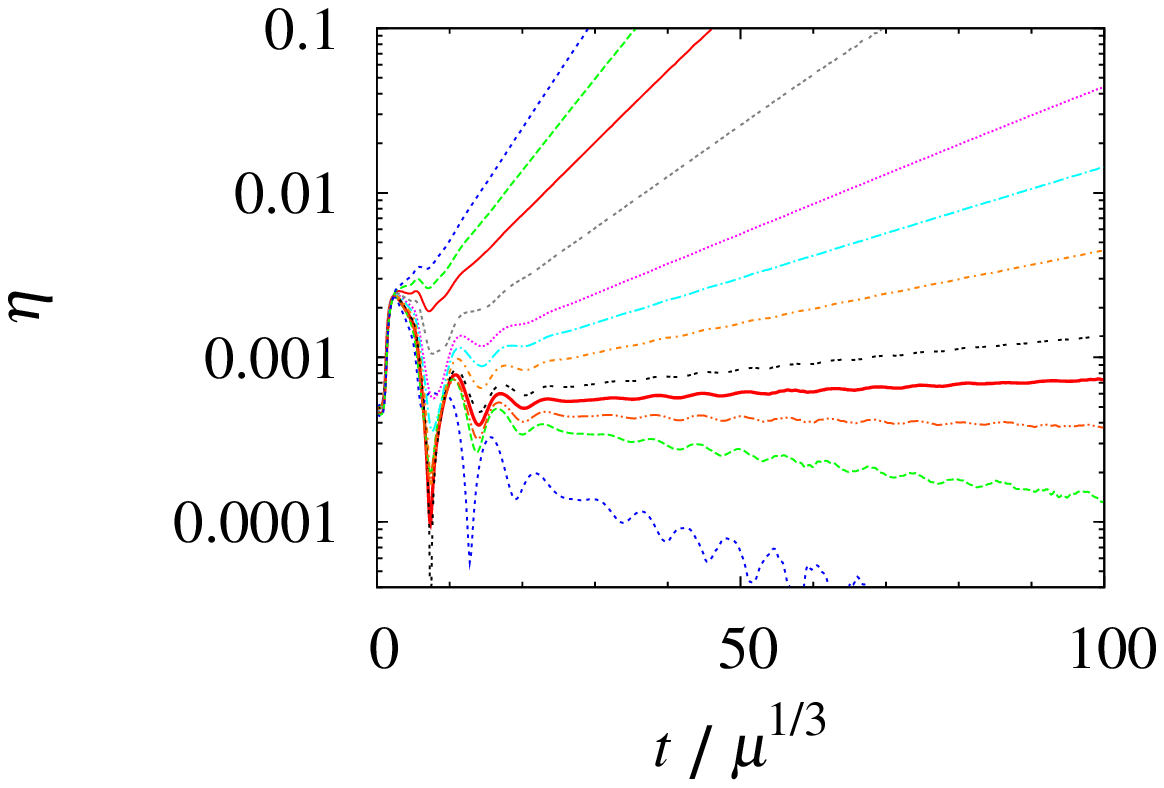}
\epsfxsize=3.2in
\leavevmode
~~(b)\epsffile{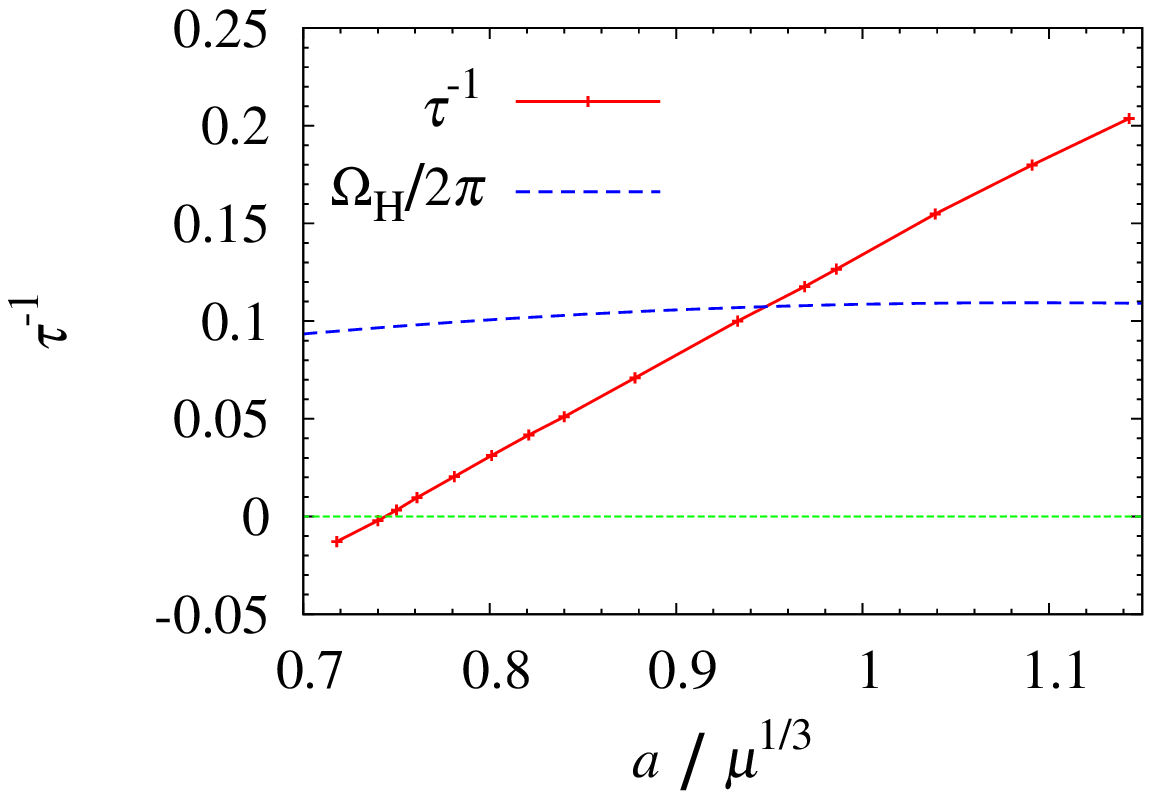}
\vspace{-3mm}
\caption{(a) Evolution of deformation parameter $\eta$ for $d=6$ and
  for the initial spin $q_i=a/\mu^{1/3} \approx 1.039$, 0.986, 0.933,
  0.878, 0.821, 0.801, 0.781, 0.761, 0.750, 0.740, 0.718, and 0.674
  (from the upper to lower curves) with $A=0.005$. (b) The growth rate
  of $\eta$, $1/\tau$, in units of $\mu^{-1/3}$ as a function of $q$
  (solid curve). The dashed curve denotes $\Omega_{\rm H}/2\pi$. For
  $q_i \agt 0.75$, the value of $\eta$ increases exponentially with
  time, and otherwise, an exponential damping is seen. For $q_i =
  0.750$ (thick solid curve in panel (a)) and 0.740 (below the curve
  of $q_i=0.750$), the growth and damping rates of $\eta$ are quite
  small, indicating that these BHs are close to the marginally stable
  state.
\label{FIG1}}
\end{figure*}

\subsection{Numerical results}\label{sec:nr}

\subsubsection{Critical spin for bar-mode instability}

During numerical simulation, we determine the apparent horizon and
calculate its area and circumferential radii to investigate the
properties of the unstable BHs. We also extract gravitational waves of
quadrupole mode in a local wave zone.

To determine the stability of a BH against bar-mode deformation, we
monitor two quantities.  One is a deformation parameter of the BH
horizon. To define this parameter, we first calculate circumferential
radii of the apparent horizon along several meridians. Specifically,
we measure the proper length of the meridians for $\varphi=0$ (and
$\pi$), $\pi/4$ (and $5\pi/4$), $\pi/2$ (and $3\pi/2$), and $3\pi/4$
(and $7\pi/4$); the proper length of each meridian for a given value
of $\varphi (<\pi)$ is defined by 
\beqn 
l_{\varphi}=2\int_0^{\pi/2} 
\Big(\sqrt{\gamma_{\theta\theta}(\varphi)}
+\sqrt{\gamma_{\theta\theta}(\varphi+\pi)} \Big) d\theta.  
\eeqn 
Then, we define a bar-mode deformation parameter 
\beq 
\eta:= {2\sqrt{(l_0-l_{\pi/2})^2+(l_{\pi/4}-l_{3\pi/4})^2} \over
  l_0+l_{\pi/2}}.  
\eeq
The value of $\eta$ is zero for an axisymmetric apparent horizon, and
increases as the deviation from the axial symmetry is enhanced. 
Thus, $\eta$ is an indicator of nonaxisymmetric bar-mode deformation.  A
drawback of the diagnostic with this quantity is that the apparent
horizon is a coordinate-dependent notion, and thus, $\eta$ may not
exactly trace the deformation of the BH, although the apparent horizon
agrees with the event horizon for a stationary spacetime, and hence,
for $\eta \ll 1$, they are likely to agree at least approximately each
other.

The other more physical quantity for measuring the deformation 
of a BH is gravitational waveform observed in a wave zone. 
This quantity should be coordinate-invariant, and thus, tells us 
whether the BH spacetime is stable or not with no ambiguity. 
In this work, we monitor a dimensionless form of gravitational waves
\beqn
&&h_+ := {\tilde \gamma_{xx}-\tilde \gamma_{yy} \over 2} 
\left[\frac{r}{\mu^{1/(d-3)}}\right]^{(d-2)/2}\\
&&h_{\times} := \tilde \gamma_{xy}
\left[\frac{r}{\mu^{1/(d-3)}}\right]^{(d-2)/2}
\eeqn
along the $z$-axis. Here, $r$ is the coordinate distance from the center, 
and $h_+$ and $h_{\times}$ are regarded as 
the plus ($+$) and cross ($\times$) modes 
of quadrupole gravitational waves in the wave zone. 

\begin{figure*}[t]
\epsfxsize=3.2in
\leavevmode
(a)\epsffile{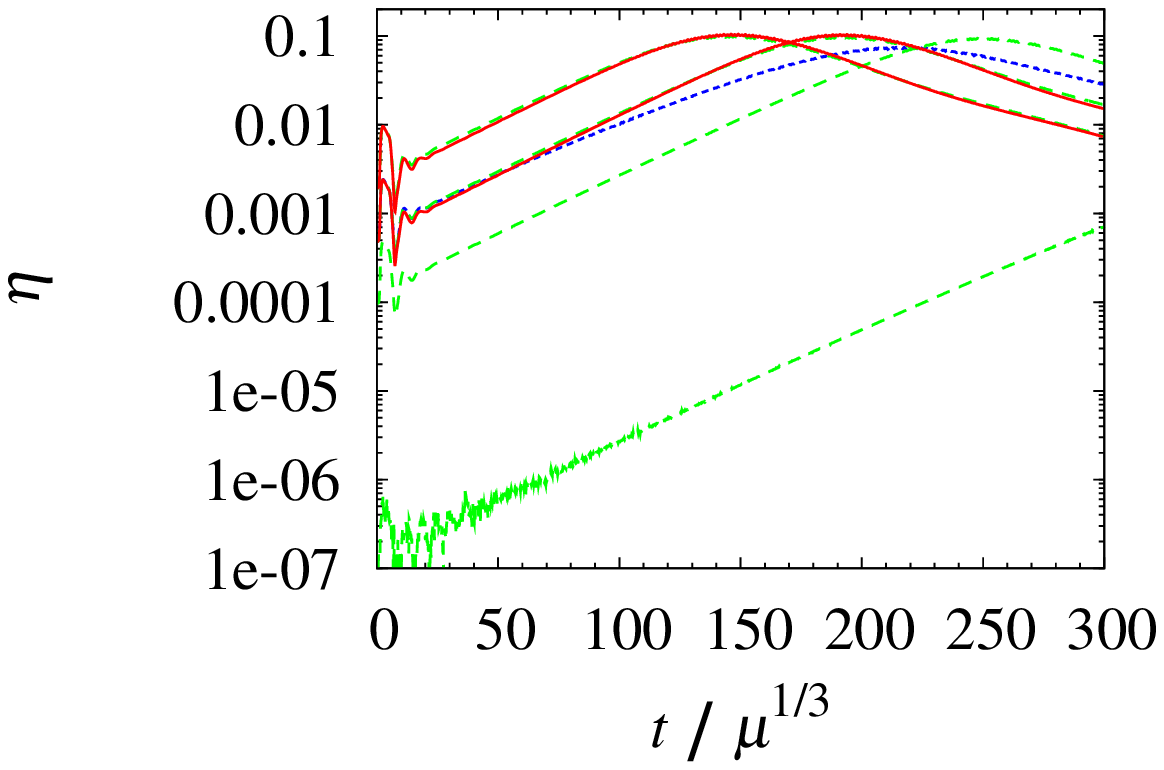}
\epsfxsize=3.2in
\leavevmode
~~~(b)\epsffile{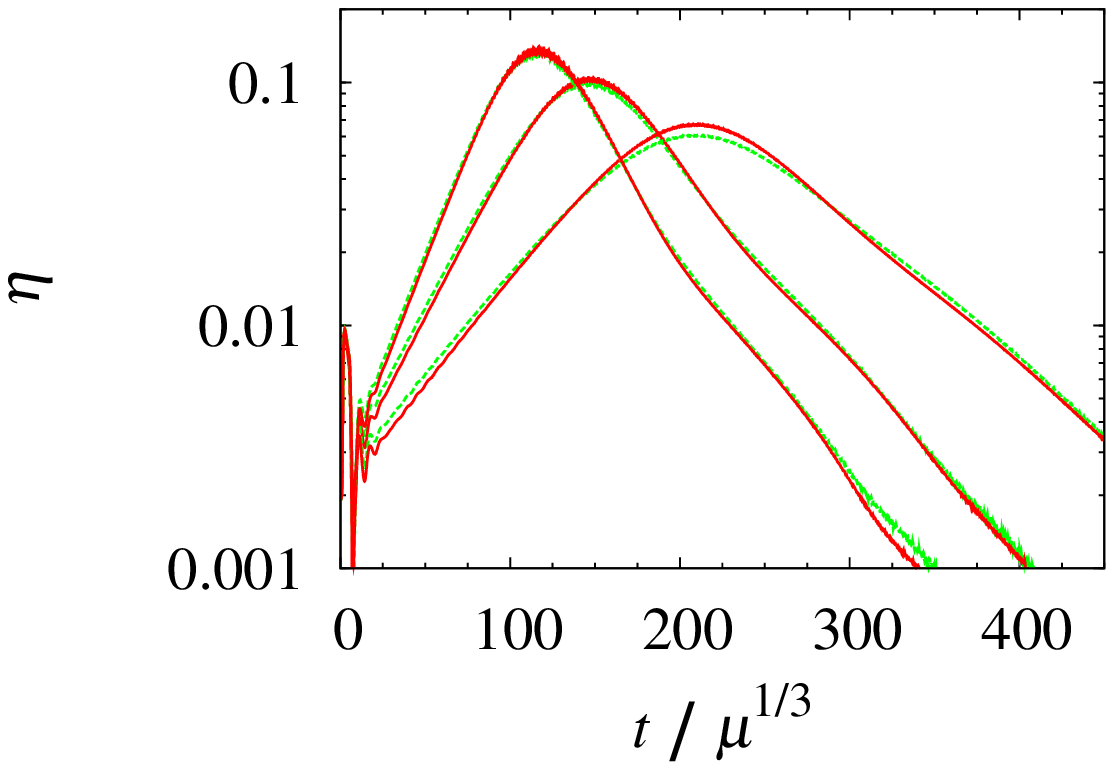} \\
\epsfxsize=3.2in
\leavevmode
(c)\epsffile{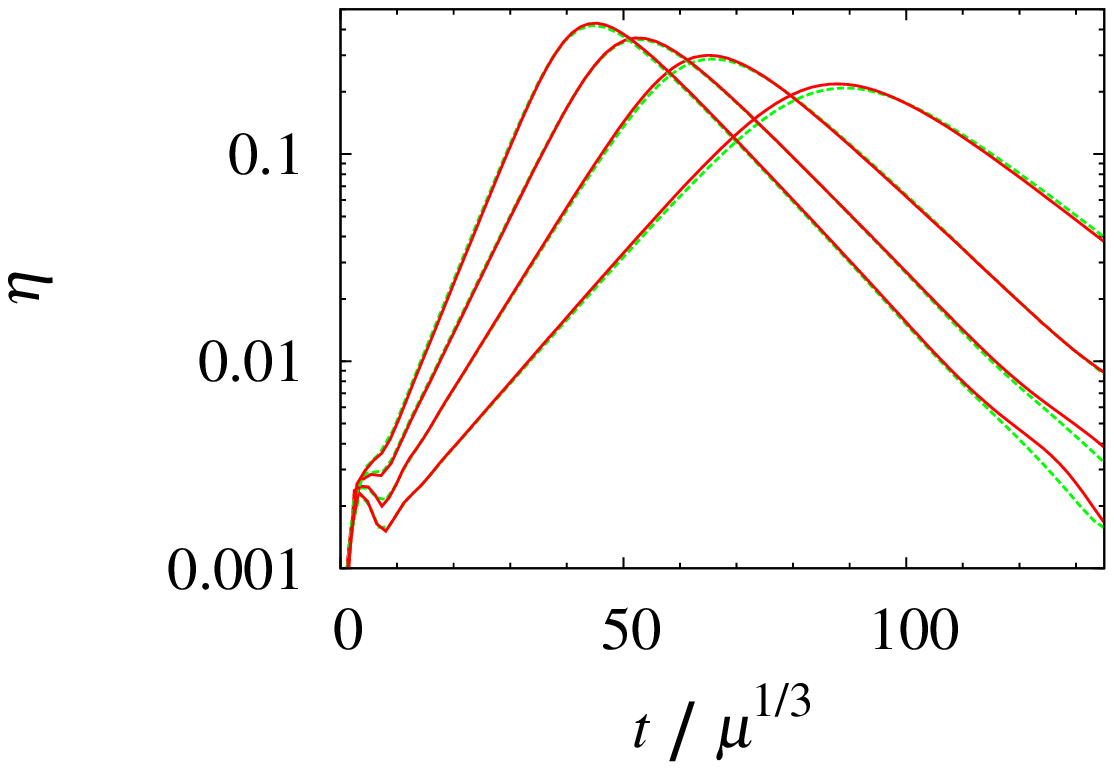}
\epsfxsize=3.2in
\leavevmode
~~~(d)\epsffile{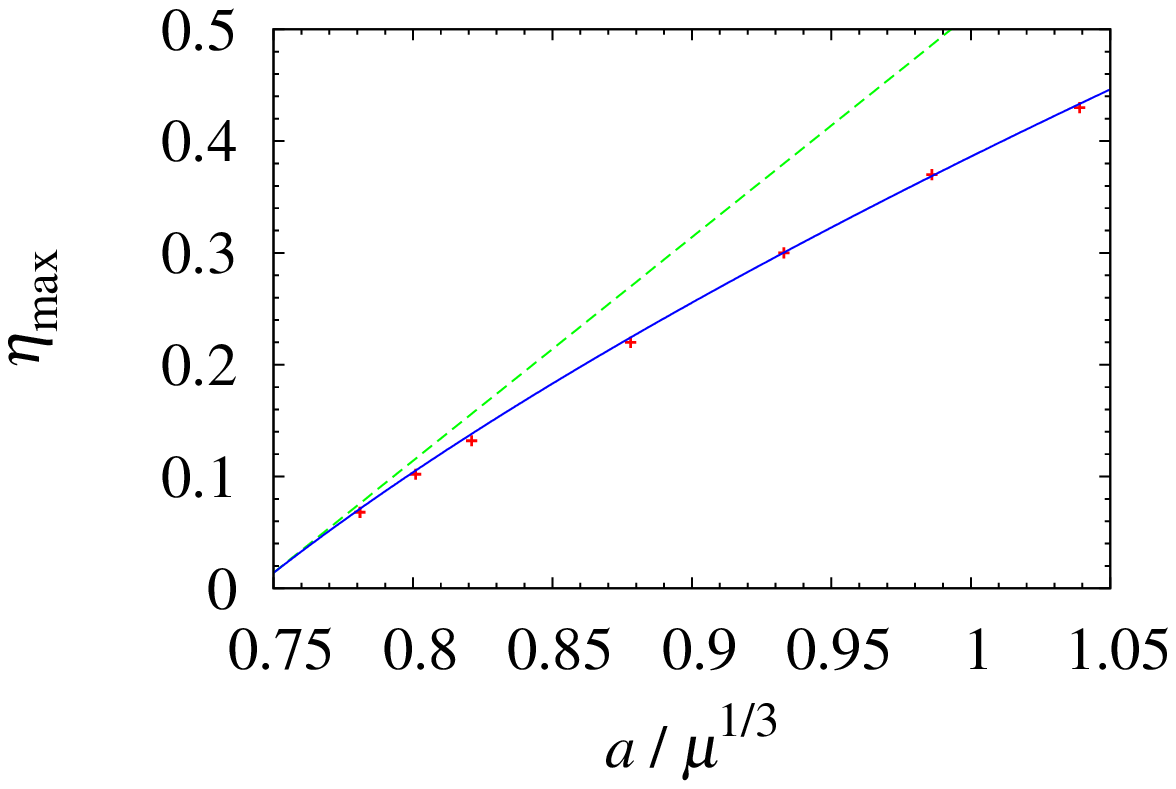}
\vspace{-3mm}
\caption{(a) Evolution of deformation parameter $\eta$ for $d=6$ and
  for $q_i=0.801$ with initial perturbation amplitude $A=0.02$, 0.005,
  0.001, and $10^{-6}$ (from the upper to lower curves) and with
  $N=30$ (dotted curve), 40 (dashed curves), and 50 (solid curves),
  respectively. (b) The same as (a) but for $q_i=0.821$, 0.801, and
  0.780 (from left to right) with $A=0.02$ and with $N=40$ (dashed
  curves) and 50 (solid curves).  (c) The same as (b) but for
  $q_i=1.039$, 0.986, 0.933, and 0.878 (from the upper to lower
  curves) with $A=0.005$.  (d) The maximum value of $\eta$ as a
  function of $q_i$.  The dashed line and solid curve denote
  $\eta_{\rm max}=2(q_i - q_{\rm crit})$ and relation (\ref{etamaxf}),
  respectively (see Sec.~VI~B for an approximate derivation of these
  relations).
\label{FIG2}}
\end{figure*}

\begin{figure*}[t]
\epsfxsize=3.2in
\leavevmode
(a)\epsffile{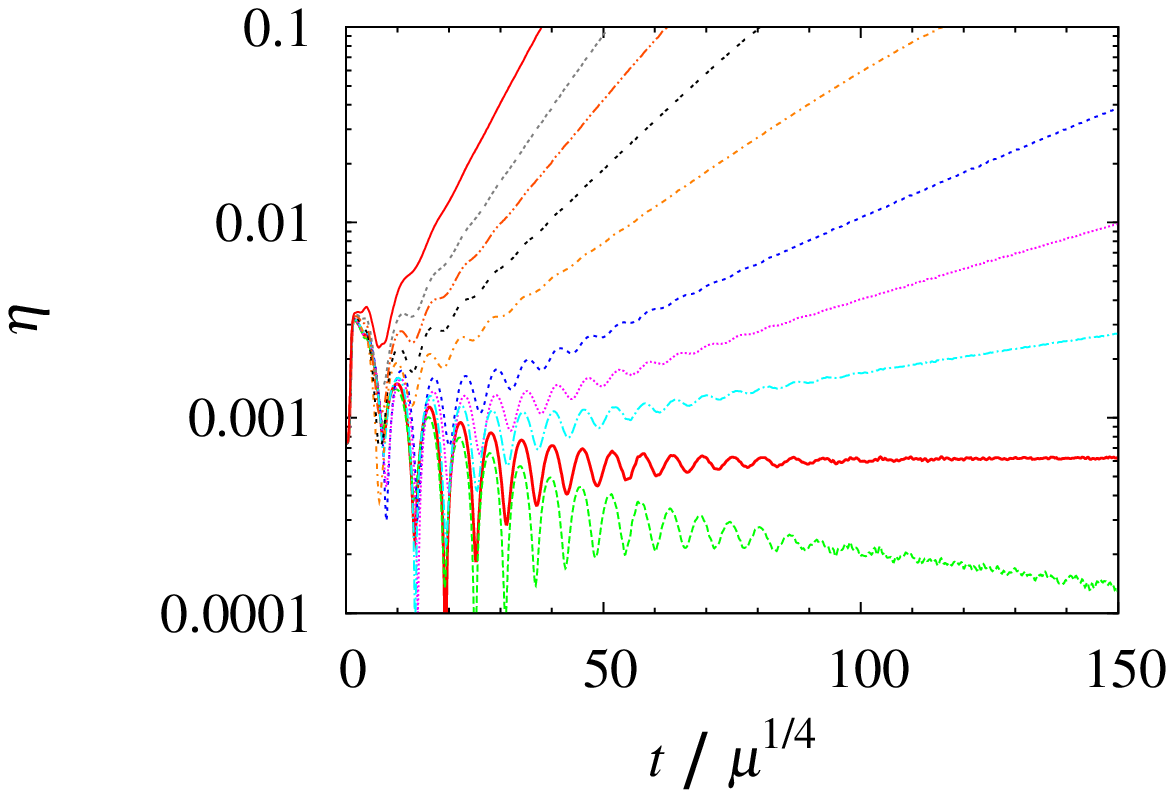}
\epsfxsize=3.2in
\leavevmode
~~~(b)\epsffile{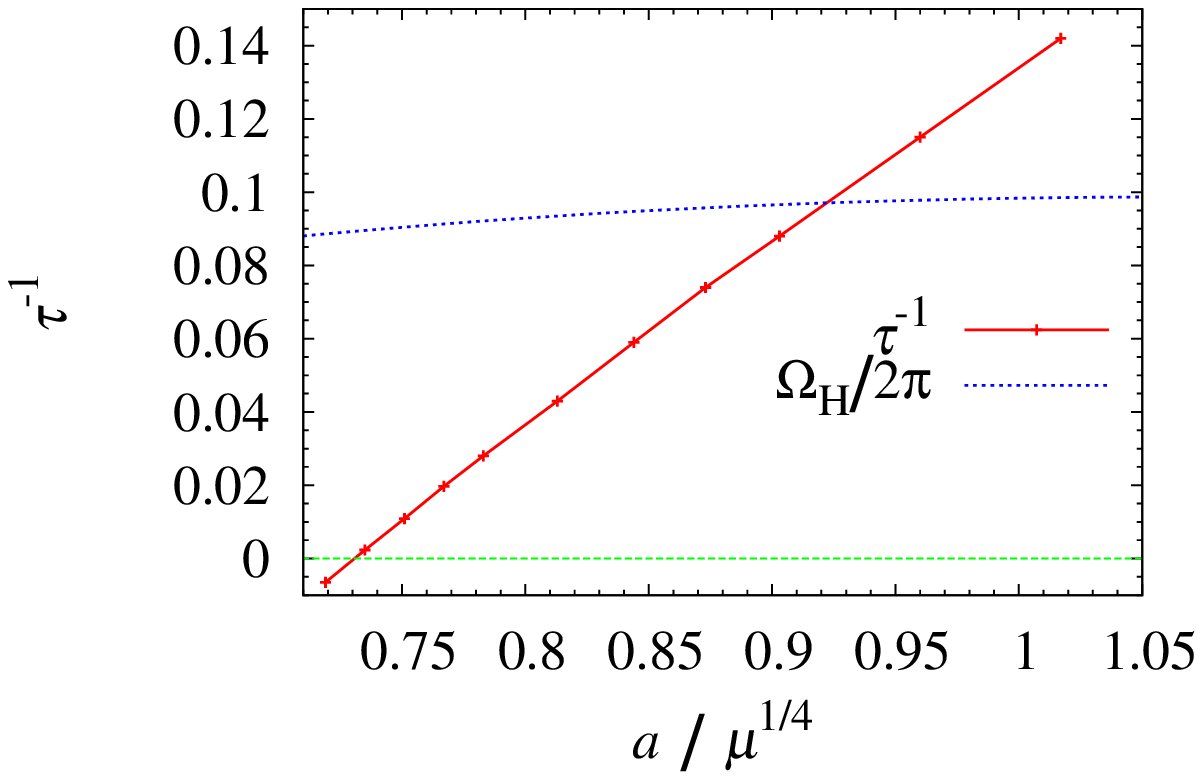} \\
\epsfxsize=3.2in
\leavevmode
(c)\epsffile{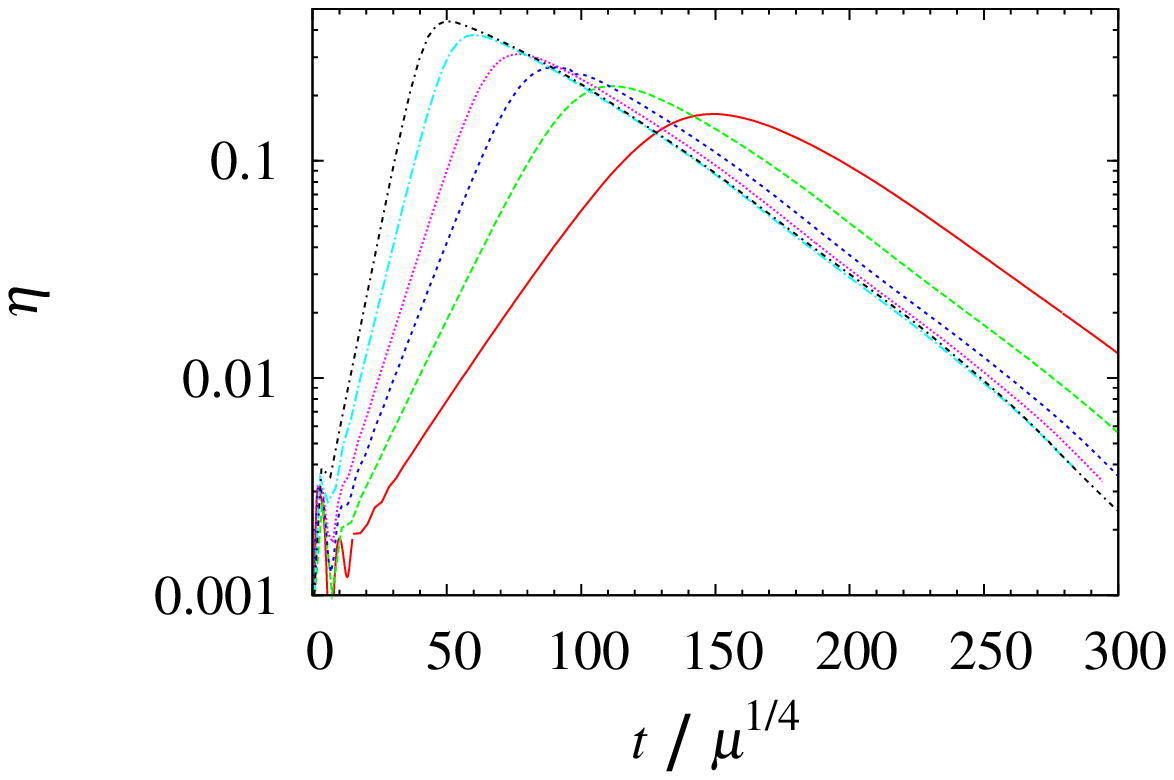}
\vspace{-3mm}
\caption{(a) The same as Fig.~\ref{FIG1} but for $d=7$ and for 
  $q_i=a/\mu^{1/4}=0.960$, 0.903, 0.844, 0.813, 0.783, 0.767, 0.751,
  0.735, and 0.719 (from the upper to lower curves) with $N=50$. For
  $q_i \agt 0.73$, the value of $\eta$ increases exponentially with
  time, otherwise, an exponential damping is seen.  For $q=0.735$, the
  growth (or damping) rate of $\eta$ is close to zero, implying that
  this BH is close to the marginally stable one.  (b) The growth rate
  of $\eta$, $1/\tau$, in units of $\mu^{-1/4}$ as a function of $q$
  (solid curve). The dashed curve denotes $\Omega_{\rm H}/2\pi$. 
  (c) The same as panel (a) but for long runs with 
  $q_i=1.017$, 0.960, 0.903, 0.873, 0.844, and 0.813. 
\label{FIG3}}
\end{figure*}

Figure \ref{FIG1}(a) plots the evolution of $\eta$ for $d=6$ and for
the initial spin $q_i=a_i/\mu^{1/3}=0.674$ -- 1.039 with
$A=0.005$. This shows that for $t/\mu^{1/3} \agt 10$, the value of
$\eta$ increases or decreases in an exponential manner as
\beqn
\eta = \eta_0 e^{t/\tau},
\eeqn
where $\tau$ is the growth timescale and $\eta_0$ is a constant.
Figure~\ref{FIG1}(b) plots the growth rate, $\tau^{-1}$, in units of
$\mu^{-1/3}$ as a function of $q_i$. We determine $\tau$ by the
least-square fitting of the curve of $\ln \eta$ with linear lines for
$15\leq t/ \mu^{1/3} \leq 100$ or for the data of $15\leq t/
\mu^{1/3}$ and $\eta \leq 0.1$.  These figures show that for $q_i \agt
0.75$, $\eta$ grows exponentially with time, and otherwise, it damps
exponentially. Therefore, the BHs with $q_i \agt 0.75$ are dynamically
unstable against bar-mode deformation [the critical value, $q_{\rm
    crit}$, is found to be $\approx 0.743$ by interpolation;
  cf. Fig.~\ref{FIG1}(b)]. Because they are dynamically unstable, such
BHs have a quasinormal mode with a negative value of the imaginary
part of eigen angular frequency, $\omega_I:=\mathrm{Im}(\omega_{\rm
  QN})$, and the bar-mode deformation grows in proportional to $e^{i
  \omega_{\rm QN} t} \propto e^{-\omega_I t}$.  On the other hand, the
BH with $q_i \alt 0.74$ is stable, and thus, $\omega_I$ are positive
for any mode (strictly speaking, there might exist a mode of $\omega_I
< 0 $ with $|\omega_I|$ less than $\sim 0.01 \mu^{-1/3}$ because we
performed the simulations only for a finite duration of order $100
\mu^{1/3}$).

Figure \ref{FIG1}(b) shows that the growth rate of the instability,
$\tau^{-1}$, increases monotonically with the spin $q$, and approximately
linearly in $q$ for $0.74 \leq q \alt 1$.  The growth rate 
($\tau^{-1}$ or $\omega_I$) is approximately described as
$C_{\tau}(q-q_{\rm crit})$ where $C_{\tau}$ is a constant $\approx
0.51 \mu^{-1/3}$.

\begin{figure*}[t]
\epsfxsize=3.2in
\leavevmode
(a)\epsffile{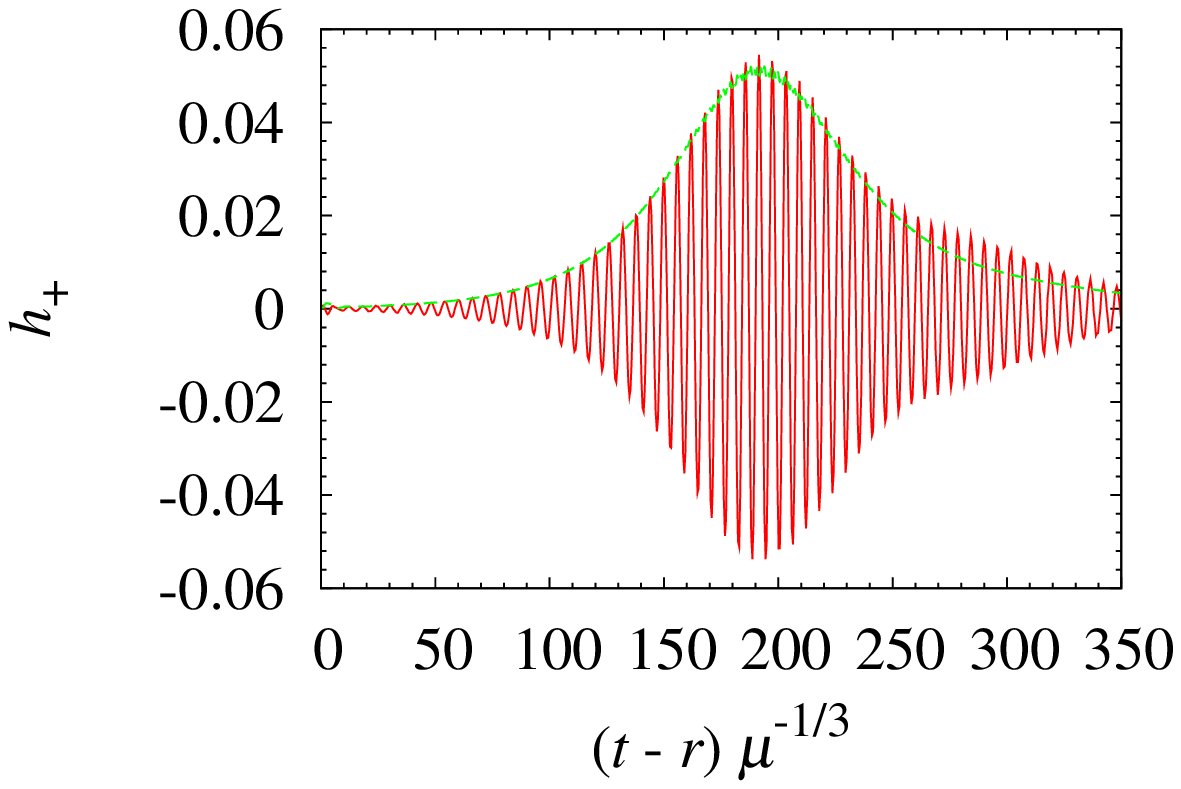}
\epsfxsize=3.2in
\leavevmode
~~~(b)\epsffile{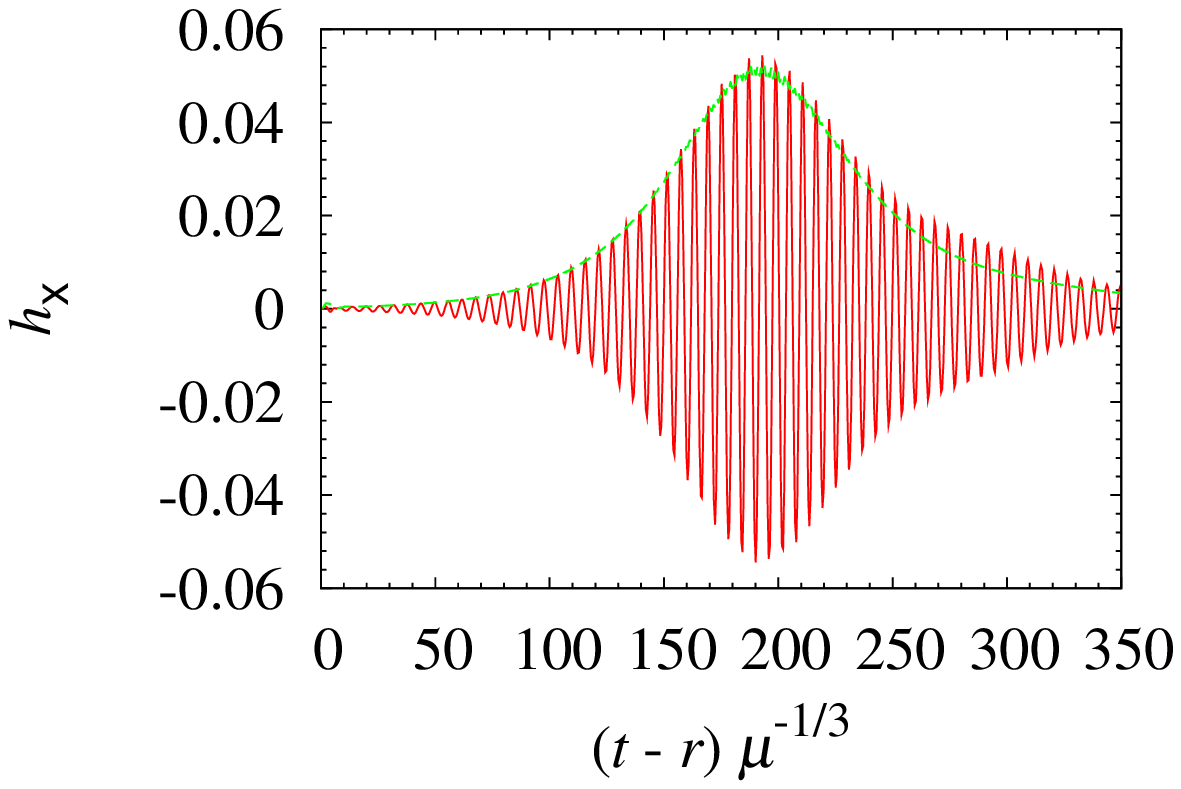}\\
\epsfxsize=3.2in
\leavevmode
(c)\epsffile{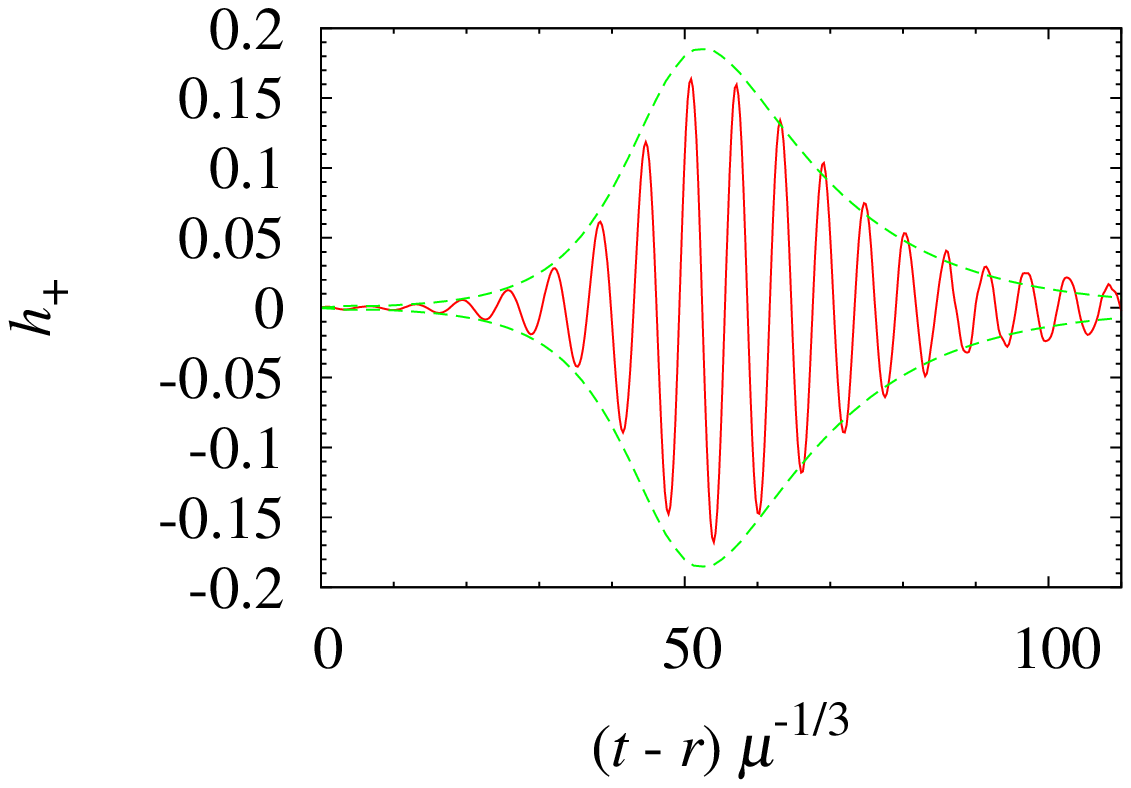}
\epsfxsize=3.2in
\leavevmode
~~~(d)\epsffile{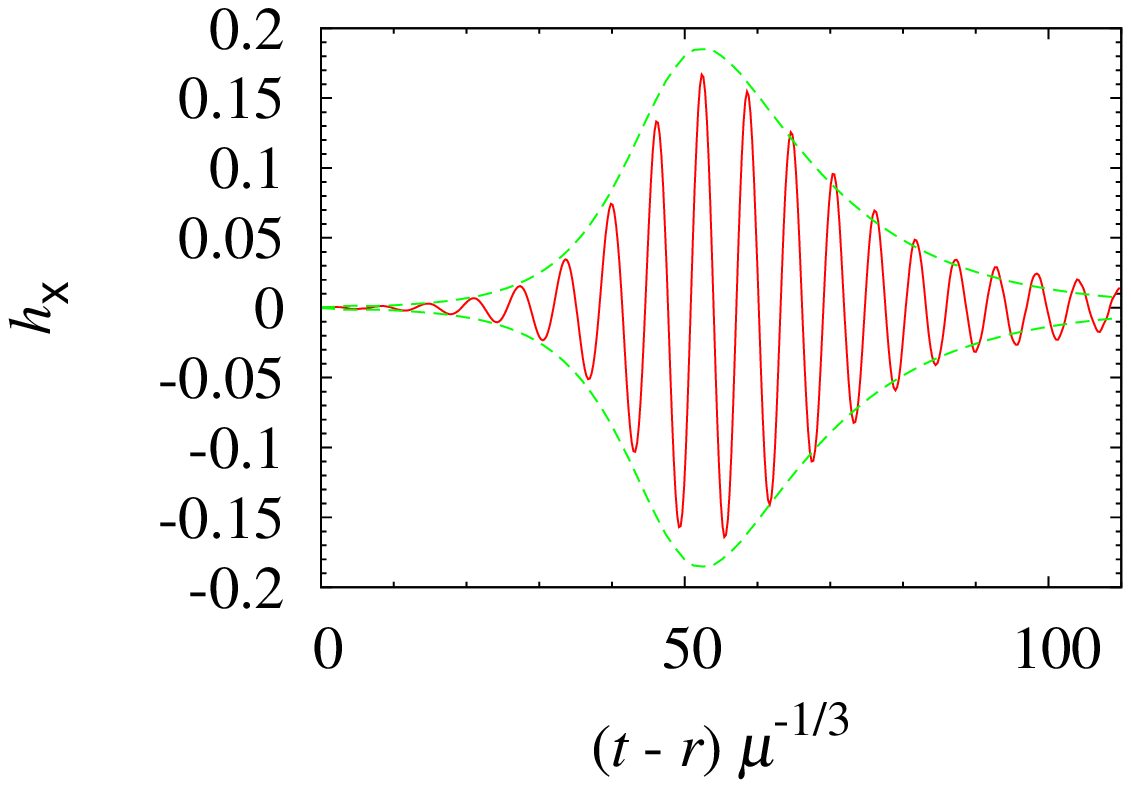}
\vspace{-3mm}
\caption{(a) and (b): $+$ and $\times$ modes of gravitational waveform
  (solid curve) from an unstable BH for $d=6$ and for $q_i = 0.801$ as
  a function of a retarded time defined by $t-r$ where $r$ is the
  coordinate distance from the center. We also plot $\eta/2$ as a
  function of $t$ (dashed curve).  The initial condition is $A=0.005$
  and the result for the grid resolution of $N=50$ is plotted.  (c)
  and (d): The same as (a) and (b) but for $d=6$ and for $q_i=0.986$
  with $A=0.005$.
\label{FIG4}}
\end{figure*}

The growth timescale of the bar-mode deformation for the BHs of spin
slightly larger than $q_{\rm crit}$ is of order $100 \mu^{1/3}$ (the
damping timescale for a stable BH with $q \alt q_{\rm crit}$ is also
of this order). This is much longer than the spin period of the BHs
defined by $2\pi/\Omega_{\rm H}=O(10\mu^{1/3})$ where $\Omega_{\rm
  H}=a/(r_+^2 + a^2)$ is the angular velocity of the BH horizon. Thus,
the instability grows slowly with a timescale much longer than the
dynamical timescale $\sim r_+/c$ for $q \sim q_{\rm crit}$.  For a
much larger value of $q$, however, the growth timescale is much
shorter: We find that $\tau$ is as short as $2\pi/\Omega_{\rm H}$ for
$q \approx 0.95$, and as $\pi/\Omega_{\rm H}$ for $q \approx 1.15$.
Simply extrapolating the approximately linear relation for
$\tau^{-1}(q)$, $\tau$ is likely to be much shorter than
$2\pi/\Omega_{\rm H}$ for $q \gg 1$. Implications of these properties
for the evolution of rapidly spinning BHs will be discussed in
Secs.~V~B and VI.

Figure \ref{FIG2}(a) plots the longterm evolution of $\eta$ for $d=6$
and for $q_i=0.801$ with several magnitudes of initial perturbation
($A=10^{-6}$, 0.001, 0.005, and 0.02) and for different grid
resolutions ($N=30$, 40, and 50 for $A=0.005$, and $N=40$ and 50 for
$A=0.02$).  This shows that the growth rate is independent of the
initial perturbation magnitude, implying that the instability sets in
irrespective of initial perturbation; even from an infinitesimally
small perturbation, the instability grows spontaneously.  The growth
rate depends weakly on the grid resolution; for a poor grid resolution
($N=30$), the growth rate is underestimated. However, for $N \geq 40$,
it appears to approximately converge, and thus, the simulations with
$N \geq 40$ (grid spacing smaller than $0.03 \mu^{1/3}$) 
have an acceptable resolution.

After the value of $\eta$ reaches the maximum, the growth of the
instability terminates, and then, $\eta$ damps exponentially with
time.  The reason for this is that during the nonlinear growth of the
bar-mode deformation, gravitational-wave emission is enhanced, and
energy and angular momentum of the BHs are carried away. As a result,
the BH spin parameter, $q$, {\em decreases} and at the saturation, it
reaches a stable state with $q < q_{\rm crit}$, for which the sign of
the imaginary part of eigen angular frequency for the corresponding
quasinormal mode changes to be positive.  The evolution of the BH
spin will be analyzed in detail in Sec.~\ref{sec4c}. Here, we show an
indirect evidence that this interpretation is correct: The saturation
value of $\eta$ is larger for the BHs of larger initial spin, because
the required spin-down fraction is $q_i-q_{\rm crit}$.
Figure~\ref{FIG2}(b) plots the evolution of $\eta$ for $A=0.02$ and
for relatively small initial spins, $q_i=0.821$, 0.801, and 0.781.
Figure~\ref{FIG2}(c) is the same as Fig.~\ref{FIG2}(b) but for large
initial spins $q_i=0.878$, 0.933, 0.986, and 1.039 with $A=0.005$.
These clearly illustrate that for the larger initial spin, a BH has to
emit more gravitational waves for spinning down to reach a stable BH,
and hence, the nonlinear growth of the bar-mode deformation has to
continue until a high saturation value of $\eta$ is reached.
Figure~\ref{FIG2}(d) plots the maximum value of $\eta$, $\eta_{\rm
  max}$, as a function of $q_i$. This clearly shows that the value of
$\eta_{\rm max}$ increases systematically with $q_i$: The relation
between $\eta_{\rm max}$ and $q_i$ is approximately written in the
form (\ref{etamaxf}) (see Sec.~VI~B for an approximate derivation of
this relation).  This result suggests that for $q_i \gg 1$, the BH may
reach a highly deformed state with $\eta \sim 1$ (although we were not
able to evolve such ultra spinning BHs in the present work.)

The numerical results obtained in this paper are approximately derived
using a semi-analytic calculation, as illustrated here.  Based on the
analytic calculations, we will show in Sec.~\ref{sec:mec} that the BH
spin should be decreased by gravitational radiation reaction. 

It is worthy to note that for the larger initial spin $q_i \sim 1$,
the damping timescale for the bar-mode deformation after the
saturation is reached depends weakly on the initial spin: The timescale 
(the duration until $\eta \leq 10^{-3}$ is reached) is $\sim
100\mu^{1/3}$.  This seems to be due to the fact that the damping
timescale depends primarily on the BH state after the saturation is
reached; i.e., the property of a BH of spin $q \alt q_{\rm crit}$
determines the damping timescale. The damping time duration is fairly long
$\sim 100\mu^{1/3}$, implying that the unstable BHs always have to
spend a long time until they reach a stationary, stable state. 

Figure \ref{FIG3}(a) plots the evolution of $\eta$ for $d=7$ and
$q_i=a/\mu^{1/4}=0.719$ -- 0.960. This shows that for $q_i \agt
0.735$, $\eta$ grows exponentially with time, and otherwise, it damps
exponentially. Figure \ref{FIG3}(b) displays the growth rate,
$\tau^{-1}$, in units of $\mu^{-1/4}$, and shows that it monotonically
increases, approximately linearly with $q$ for $0.719 \leq q \alt 1$
as in the case of $d=6$.  The critical spin for the onset of the
bar-mode instability is determined as $q_{\rm crit} \approx 0.730$,
and near $q=q_{\rm crit}$ the growth rate ($\tau^{-1}$ and $\omega_I$)
behaves as $C_{\tau}(q-q_{\rm crit})$ where $C_{\tau}$ is a constant
$\sim 0.54 \mu^{-1/4}$.  It is interesting to note that the
coefficient $C_{\tau} \mu^{1/(d-3)}$ for $d=7$ is close to that for
$d=6$.

The critical spin, $q_{\rm crit}$, for the onset of the bar-mode
instability is also very close to that for $d=6$. Simulations for
$d=8$ also clarified that the critical spin is $q_{\rm crit}\approx 0.77$,
where $q=a/\mu^{1/5}$ for $d=8$, again close to those for $d=6$ and
$7$. This suggests that the critical value of $q$ depends weakly on
the value of $d$ as long as $d \geq 6$ (note, however, that for $d=5$,
we found it a relatively large value, $\approx 0.87$ \cite{SY09}).
The values of $q_{\rm crit}$ and the corresponding values of
$a_*=a/r_+$ are summarized in Table \ref{table0}.

\begin{table}[tb]
\caption{The values of $q_{\rm crit}$ and the corresponding values 
of $a_*$ and $C_p/C_e$ for $d=5$ -- 8. $C_p/C_e$ is defined in 
Eq.~(\ref{cpce}).
}
\begin{ruledtabular}
\begin{tabular}{c|cccc}
 $d$ &  $5$ &$6$ & $7$ & $8$  \\
  \hline
 $q_{\rm crit}$ & $0.87$ &$0.74$ & $0.73$ & $0.77$ \\
 $a_*$         & $1.76$ &$0.91$ & $0.83$ & $0.86$ \\
 $C_p/C_e$     & $0.38$ &$0.65$ & $0.68$ & $0.67$ \\
  \end{tabular}
\end{ruledtabular}
\label{table0}
\end{table}

As in the case $d=6$, the growth rate, $\tau^{-1}$, increases
monotonically with the value of $q_i$ for $d=7$. For $q_i \agt 0.92$,
it is larger than the BH spin frequency $\Omega_{\rm H}/2\pi$. This
indicates the growth timescale is much shorter than the BH spin for
$q_i \gg 1$ irrespective of the dimensionality.

Figure \ref{FIG3}(c) plots the longterm evolution of $\eta$ for $d=7$ with
higher spins, $q_i=0.813$ -- 1.017. For these BHs, the growth of the
bar-mode deformation saturates at a large value of $\eta$ as
$\eta_{\rm max}=0.1$--0.5 due to gravitational radiation reaction, and
then, $\eta$ decreases exponentially as in the case of $d=6$.  At the
saturation, the BH spin parameter, $q$, is likely to be slightly
smaller than $q_{\rm crit}$.  As in the case of $d=6$, the damping
rate of the bar-mode deformation after the saturation is reached
depends only weakly on the initial value of $q_i$.  The reason for
this is that it depends primarily on the spin achieved just after the
saturation is reached ($q \alt q_{\rm crit}$), as mentioned before.
One point to be noted is that the damping timescale is much longer
than the growth timescale for $q_i \sim 1$ (in the 6D case, two
timescales are not very different and smaller than $50
\mu^{1/3}$). The damping timescale is $\agt 100 \mu^{1/4}$ in the 7D
case. This suggests that it always takes a long time for the unstable
BHs to reach a stable state for $d \geq 7$.


To confirm that the bar-mode deformation of the BHs is indeed physical
(the growth of $\eta$ is not due to a coordinate choice for finding
apparent horizons), we extract gravitational waves in the wave zone.
Figure \ref{FIG4} plots gravitational waveforms, $h_+$ and
$h_{\times}$, as a function of a retarded time $t-r$ for $d=6$ and for
$q_i \approx 0.801$ and 0.986. We note that gravitational waveforms
for different values of $q_i$ are qualitatively similar.  Here, $r$
denotes the coordinate distance of extracting gravitational waves. We
also plot together evolution of $\eta/2$ as a function of $t$. As in
the behavior of $\eta$, the amplitude of gravitational waves increases
exponentially with time in the early phase, and after the saturation
point is reached, it starts exponential damping. For the larger
initial spin, the growth timescale is shorter; for $q_i=0.986$, the
grow timescale is comparable to the oscillation period.  $h_+$ and
$h_{\times}$ behave essentially in the same manner except for a phase
difference of $\pi/2$. The amplitude of gravitational waves is
approximately equal to $\eta/2$ for the small amplitude and slightly
smaller than $\eta/2$ for the amplitude $(h_+^2+h_{\times}^2)^{1/2}
\agt 0.1$. This slight disagreement in the amplitude for the
high-amplitude case is likely due to the fact that the nonlinear
deformation of the BH enhances the amplitude of gravitational waves of
modes other than the quadrupole ones, suppressing the quadrupole
modes. It is worthy to point out that the oscillation frequency
remains approximately constant (besides a small secular shift
associated with the evolution of the BHs). This indicates that one
fundamental quasinormal mode contributes to the instability.

\subsubsection{Evolution of spin and area}\label{sec4c}

\begin{figure}[th]
\epsfxsize=3.3in
\leavevmode
\epsffile{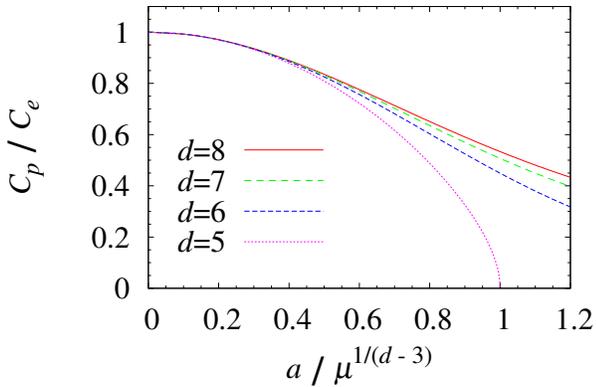}
\vspace{-3mm}
\caption{$C_p/C_e$ as a function of $q=a/\mu^{1/(d-3)}$ for $d=5$ -- 8.  
\label{FIG5}}
\end{figure}

\begin{figure*}[t]
\epsfxsize=3.3in
\leavevmode
(a)\epsffile{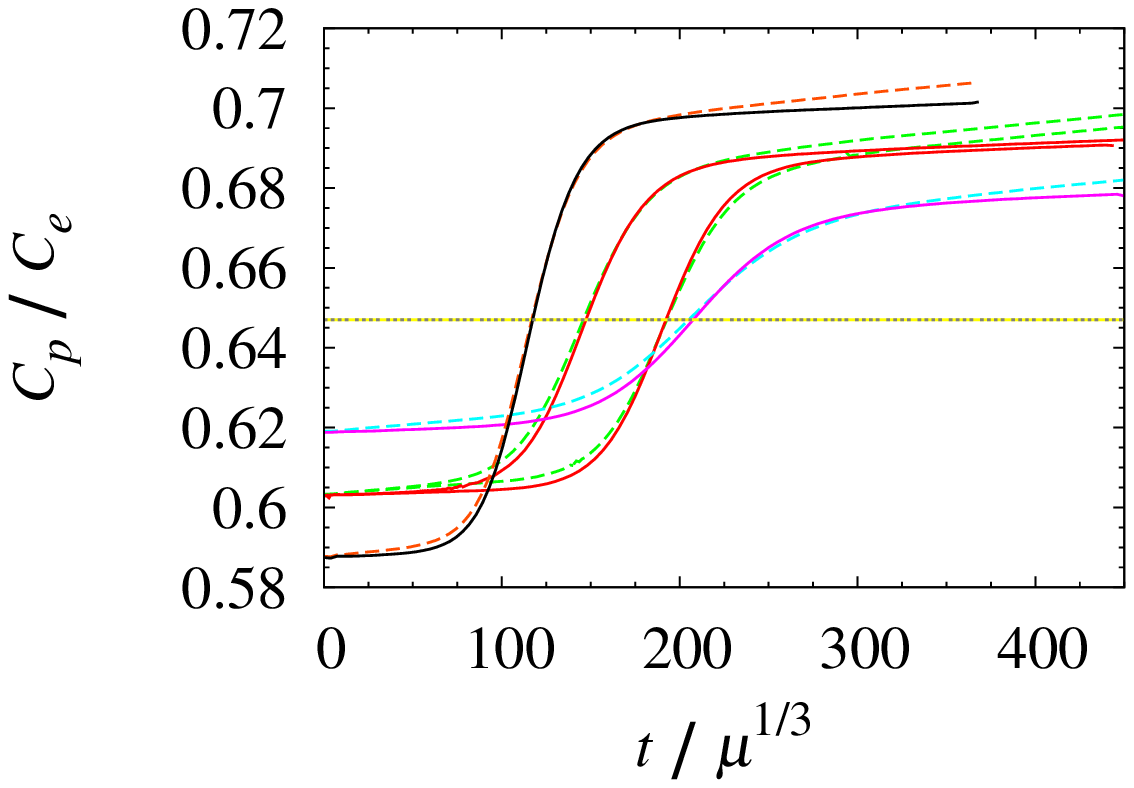}
\epsfxsize=3.3in
\leavevmode
~~(b)\epsffile{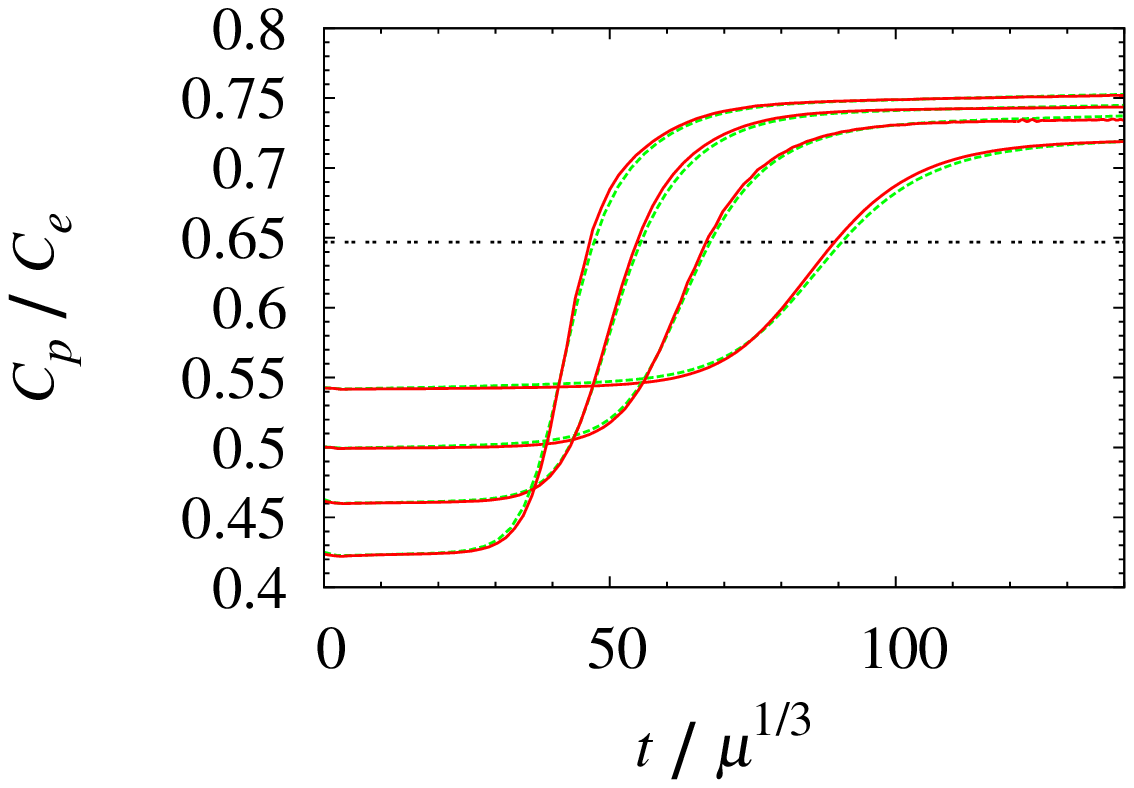}
\vspace{-3mm}
\caption{(a) Evolution of $C_p/C_e$ for $d=6$ and for relatively
  small initial spins $q_i=a/\mu^{1/3}=0.821$ ($C_p/C_e \approx 0.587$
  at $t=0$), 0.801 ($C_p/C_e \approx 0.602$ at $t=0$), and 0.781
  ($C_p/C_e \approx 0.618$ at $t=0$) as a function of time. For
  $q_i=0.801$, the results with $A=0.02$ and 0.005 are plotted.  For
  $q_i=0.821$ and 0.781, the results with $A=0.02$ are plotted. The
  solid and dashed curves denote the results for $N=50$ and 40,
  respectively.  The thin dotted line denote $C_p/C_e=0.647$ which is
  the value of $C_p/C_e$ for $q=q_{\rm crit}$. For $q_i=0.821$, we
  stopped the simulation at $t/\mu^{1/3} \approx 370$ because the BH
  reaches approximately stationary state.  (b) The same as (a) but for
  the large initial spins $q_i=0.878$, 0.933, 0.986, and 1.039 with
  $A=0.005$. $C_p/C_e \approx 0.542$, 0.499, 0.460, and 0.422 at
  $t=0$, respectively.
\label{FIG6}}
\end{figure*}

\begin{figure}[th]
\epsfxsize=3.4in
\leavevmode
\epsffile{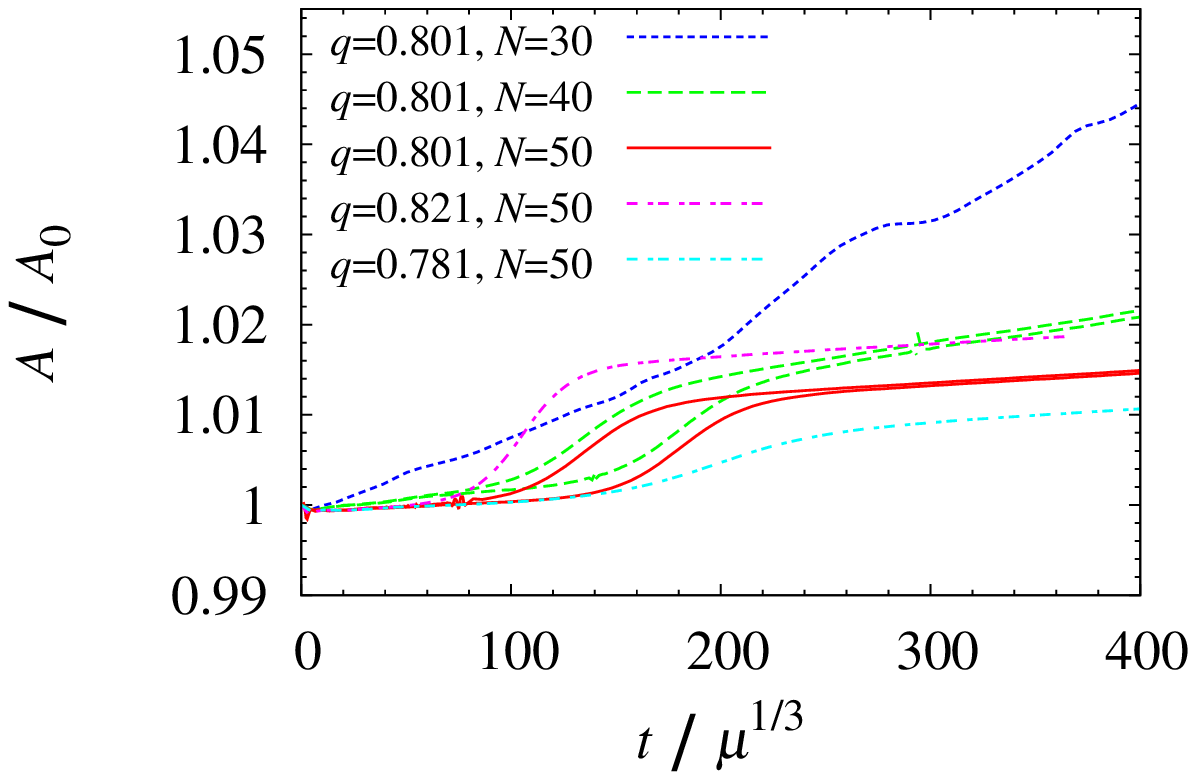}
\vspace{-5mm}
\caption{The same as Fig.~\ref{FIG6} but for the area of apparent
  horizon for $q_i=0.801$ with $N=30$ (dotted curve), 40 (dashed
curves), and 50 (solid curves) and for $q_i=0.781$ and 0.821 with
$N=50$ (lower and upper dot-dashed curves).  $A_0$ is the initial
value of the area, $\Omega_{d-2}\mu r_+$.
\label{FIG7}}
\end{figure}

Now, we illustrate how the BH spin evolves as a result of
gravitational radiation reaction.  To approximately determine the
value of the BH spin from the shape of apparent horizon, we calculate
the ratio of a polar circumferential length, $C_p$, to the
equatorial circumferential length, $C_e$, for which we define
\beqn
&&C_p:={l_0 + l_{\pi/2} \over 2},\\
&&C_e:=\int_0^{2\pi} \sqrt{\gamma_{\varphi\varphi}} d\varphi, 
\eeqn
where the integral for $C_e$ is performed along the 
surface of the apparent horizon at $\theta=\pi/2$. 
For the Myers-Perry BH of single spin, the ratio $C_p/C_e$ is given by 
\beqn
{C_p \over C_e}={2 r_+^{d-3} \over \pi \mu}
\int_0^{\pi/2} \sqrt{1 + {a^2 \cos^2\theta \over r_+^2}} d\theta,
\label{cpce}
\eeqn
which is a monotonically decreasing function of $q=a/\mu^{1/(d-3)}$
for any given number of $d$ (see Fig.~\ref{FIG5}).
Thus, as far as the state of a BH is close to an axisymmetric
stationary BH, we may use it for measuring the spin, $q$. We note that
5D BHs become dynamically unstable only for a small value of $C_p/C_e
\alt 0.38$ \cite{SY09}, whereas for $d \geq 6$, the critical values of
$C_p/C_e$ are rather large universally; $C_p/C_e=0.65$ for $d=6$, 0.68
for $d=7$, and 0.67 for $d=8$ (listed in Table I). This illustrates
that the 5D BHs are qualitatively different from the higher-dimensional 
BHs with $d \geq 6$.

Figure \ref{FIG6}(a) plots the evolution of $C_p/C_e$ for $d=6$ and
for relatively small initial spins $q_i = 0.821$, 0.801, and
0.781. The initial value of this ratio is $\approx 0.587$ for
$q_i=0.821$, $\approx 0.602$ for $q_i=0.801$, and $\approx 0.618$ for
$q_i=0.781$, respectively.  When $\eta$ is much smaller than 0.1, the
value of $C_p/C_e$ remains approximately constant. With the nonlinear
growth of the bar-mode deformation, it starts increasing due to the
decrease of the spin by gravitational radiation reaction, and it
eventually settles to a constant $\approx 0.700$ for $q_i=0.821$,
$\approx 0.690$ for $q_i=0.801$, and $\approx 0.677$ for $q_i=0.781$,
respectively, when $\eta \ll 1$ being achieved (we adopt the values of
$C_p/C_e$ when we stopped the simulations). We note that with a poor
grid resolution, the value of $C_p/C_e$ spuriously increases even for
$\eta \ll 1$ because numerical dissipation tends to decrease the BH
spin. This spurious variation is suppressed for the better grid
resolutions (see also Fig.~\ref{FIG7}).  The final values of $C_p/C_e$
indicate the final spin $q_f \approx 0.675$, 0.688, and 0.705 for
$q_i=0.821$, 0.801, and 0.781, respectively, for which the BHs appear
to be stable against any perturbation. It is worthy to note that the
larger initial spin leads to the smaller final spin because a larger
amount of gravitational-wave emission is enhanced during the evolution
(see also Sec.~\ref{sec:mec} for the reason to this); the value
of $q_f$ is approximately written as $q_f = 2q_{\rm crit}-q_i$ for
$q_i=0.781$ -- 0.821.

Figure \ref{FIG6}(b) plots the evolution of $C_p/C_e$ for $d=6$ and
for larger initial spins $q_i=0.878$, 0.933, 0.986, and 1.039. As
in the case of relatively small initial spins, the value of
$C_p/C_e$ increases with the evolution, and eventually settles to a
constant after the BH is stabilized by gravitational radiation
reaction.  However, the final value of $C_p/C_e$ depends only weakly
on the initial spin, and hence, the final spin estimated from
$C_p/C_e$ is much larger than the value of $q_f=2 q_{\rm crit}-q_i$. Figure
\ref{FIG6}(b) shows that the fractional change of $C_p/C_e$ after the
BH is stabilized is smaller than that before the stabilization.  This
indicates that the fraction of spin-down is suppressed after the
evolution of the nonlinear bar-mode deformation is saturated; again,
this seems to be due to the fact that the evolution of the BHs after
the saturation is achieved is determined by the property of the BH of
spin slightly smaller than $q_{\rm crit}$. For $q_i=0.878$ -- 1.039,
the final spin is relatively in a narrow range as $q_f \approx 0.65$
-- 0.61.  The results we found suggest that even for $q_i \gg 1$, the
final spin $q_f$ will not be close to zero, but moderately large as
$\sim 0.6$. 

Qualitatively the same results for the final spin are obtained for
$d=7$.  In this case, $C_p/C_e$ relaxes to $0.72 \pm 0.01$ for $0.85 
\alt q_i \alt 1$, implying that the final spin is in a narrow range 
$q_f=0.66$ -- 0.69; even for $q_i \gg 1$, the final spin $q_f$ will
not be close to zero, but moderately large as $\sim 0.65$. The results
of $q_f$ for $d=6$ and 7 together with the weak dependence of $q_{\rm
  crit}$ on $d$ suggest that also for $d \geq 8$, the final spin will
be moderately large $\agt 0.6$ -- 0.7.

The area of the BH has to increase as a result of spin-down
\cite{unique}.  Figure \ref{FIG7} plots the evolution of the horizon
area for $d=6$ and $q_i=0.781$, 0.801, and 0.821 with $N=50$.  For
$q_i=0.801$, we plot the results for different grid resolutions.
Figure \ref{FIG7} shows that the area indeed increases.  As noted
above, the area gradually increases with time even in the state of
$\eta \ll 1$. This is partly due to numerical dissipation. 
Indeed, the error decreases significantly with improving the
grid resolution; for $N = 50$, the spurious increase of the area seems
to be negligible, and hence, we can determine the magnitude of the
physical increase due to gravitational-wave emission.  The area
increases by $\approx 1.0$, 1.4, and 1.8\% for $q_i=0.781$, 0.801, and
0.821, respectively. For the larger initial spins, the
area increases more due to gravitational radiation reaction. A 
noteworthy fact is that the fractional increase of the area is
much small than those for spin and mass energy of the BH 
(cf. Fig.~\ref{FIG9}).

\section{Properties of unstable black holes}\label{sec:mec}

\subsection{Conditions for spontaneous gravitational wave emission}

The first law of BH thermodynamics allows us to determine the
variation in the BH area $\delta A$ as \cite{EM03}
\beq
{\kappa \over 8\pi G_d} \delta A = -\Omega_{\rm H} 
\delta J + \delta E, \label{FIRST}
\eeq
where 
$\kappa$ is the surface gravity of the BH horizon, 
$\kappa = [2r_+^{d-3}+(d-5) \mu]/(2\mu r_+)$, 
which is positive for the BHs considered in this paper. 
$\delta E$ and $\delta J$ are the variations of energy
and angular momentum of the BH, and if the energy and angular momentum 
are carried away by gravitational waves, they should be negative. 
If we assume that gravitational waves of monochromatic wavelength 
are emitted, the following relation holds; 
\beq
\delta E= {\omega \over m} \delta J < 0. \label{dedj}
\eeq
Here, $\omega$ is the (real) angular frequency of the unstable mode
[we regard it as $\omega= \mathrm{Re} (\omega_{\rm QN})$ in the
  following], and $m$ is the azimuthal quantum number for which we set
$m=2$ because we focus on the bar-mode instability.  The assumption of
monochromatic wave emission is approximately correct because
gravitational waves associated with the fundamental quasinormal mode
are likely to be most strongly emitted. Indeed, Fig.~\ref{FIG4}
indicates that this is the case.

Substituting the relation (\ref{dedj}) into Eq.~(\ref{FIRST}), we obtain
\beqn 
{\kappa \over 8\pi G_d} \delta A= 
\left( \Omega_{\rm H} - {\omega \over m} \right) 
|\delta J|. \label{cond0}
 \eeqn
This shows that only for $\Omega_{\rm H} > \omega/m$, $\delta A$
becomes positive and the evolution by emission of gravitational waves
is allowed without violating the area theorem \cite{unique}.  This is
the so-called superradiance condition \cite{TP74}: If this condition
holds, the energy flux of ingoing waves of an angular frequency 
$k=\omega-m\Omega_{\rm H}$ at the BH horizon becomes negative because
$k < 0$. 

This is a necessary condition that the unstable quasinormal mode
should satisfy for the spontaneous emission of gravitational waves.
However, this is not the sufficient condition.  The superradiance is
similar to secular instability in the terminology of an instability
for rotating stars \cite{CH69,FS78}, because for the superradiance
emission of gravitational waves from a stable BH with $\omega_I > 0$,
one needs an artificial wave injection which satisfies the condition
$k < 0$. We note that for the spontaneous emission of gravitational
waves, one further needs to require the presence of a quasinormal mode
with negative imaginary part for its angular frequency, $\omega_I <
0$.

\begin{figure*}[t]
\epsfxsize=5.in
\leavevmode
\epsffile{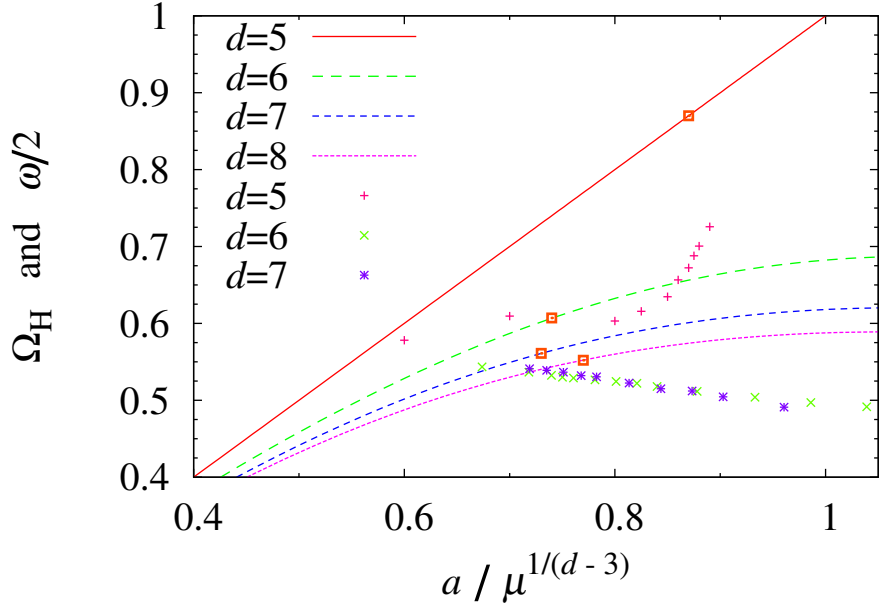}
\vspace{-5mm}
\caption{$\Omega_{\rm H}$ as a function of $q=a/\mu^{1/(d-3)}$ for
  $d=5$ -- 8 (curves), and $\omega/m=\omega/2$ for selected values of
  the spin parameter for $d=5$ -- 7 (points). The units of the vertical
  axis are $\mu^{-1/(d-3)}$. We omit the points for $d=8$ because they
  are located in the vicinity of the points for $d=6$ and 7: The
  values of $\omega \mu^{1/(d-3)}/2$ are in a narrow range 0.49 -- 0.55
  for the chosen values of $q$ and for $d=6$ -- 8. The squares on the
  curve of $\Omega_{\rm H}$ approximately denote the critical points
  for the onset of bar-mode instability (i.e., for spontaneous
  gravitational-wave emission).
\label{FIG8}} 
\end{figure*}

To confirm that the superradiance condition is satisfied for the
oscillation mode of unstable BHs, we first determine the real value of
$\omega$ by performing the Fourier transformation of $h_+$ for the
BHs.  We identify the angular frequency at the spectrum peak as
$\omega$.  Figure \ref{FIG8} plots the resulting values for $\omega/2$
(points) as well as $\Omega_{\rm H}$ (curves) as functions of $q$. We
also plot the results for 5D BHs obtained in the previous paper
\cite{SY09}.  This figure shows that the unstable modes with $\omega_I
< 0$ always satisfy the superradiance condition $\Omega_{\rm H} >
\omega/2$ and gravitational waves can be spontaneously emitted from
the unstable BHs. A noteworthy point is that
$\omega\mu^{1/(d-3)}(\approx 1.0$--1.1) depends weakly on the spin
$q$; for $d=6$, $\omega$ decreases slowly with $q$ for $0.7 \leq q
\alt 1.1$ and for $d=7$ and 8, $\omega$ behaves in a manner similar to
that of $d=6$ for $0.7 \leq q \alt 1$.

In the following, we refer to the value of $q$, at which the real part
of the quasinormal frequency becomes equal to the threshold frequency
for the superradiance, as $q_{\rm SR}$.  The superradiance condition
is satisfied for the quasinormal modes when $q\agt q_{\rm SR}$, where
$q_{\rm SR}\approx 0.6$ for $d=5$, $q_{\rm SR} \approx 0.65$ for
$d=6$, $q_{\rm SR} \approx 0.7$ for $d=7$, and $q_{\rm SR} \approx
0.75$ for $d=8$.  Because the maximum value of $\Omega_{\rm
  H}\mu^{1/(d-3)}$ decreases as the number of $d$ is increased and
$\omega \mu^{1/(d-3)}$ for a given value of $q$ depends weakly on $d$,
the value of $q_{\rm SR}$ will increase as $d$ is increased.  This
indicates that the critical value of spin, $q_{\rm crit}$, for the
onset of the bar-mode instability (i.e., for the spontaneous
gravitational-wave emission) will also increase with $d$ for 
$d \geq 8$.

We note that Fig.~\ref{FIG8} shows $\Omega_{\rm H}-\omega/m \ll
\Omega_{\rm H}$ for $6 \leq d \leq 8$, and thus, the area increases
only slightly even if a large amount of angular momentum is dissipated
by gravitational-wave emission [see Eq.~(\ref{cond0})]. This explains
why the area increases only slightly (see Fig.~\ref{FIG7}) even when
$q$ changes by a large factor $\sim 0.1$ -- 0.2.

\begin{figure*}[t]
\epsfxsize=3.3in
\leavevmode
\epsffile{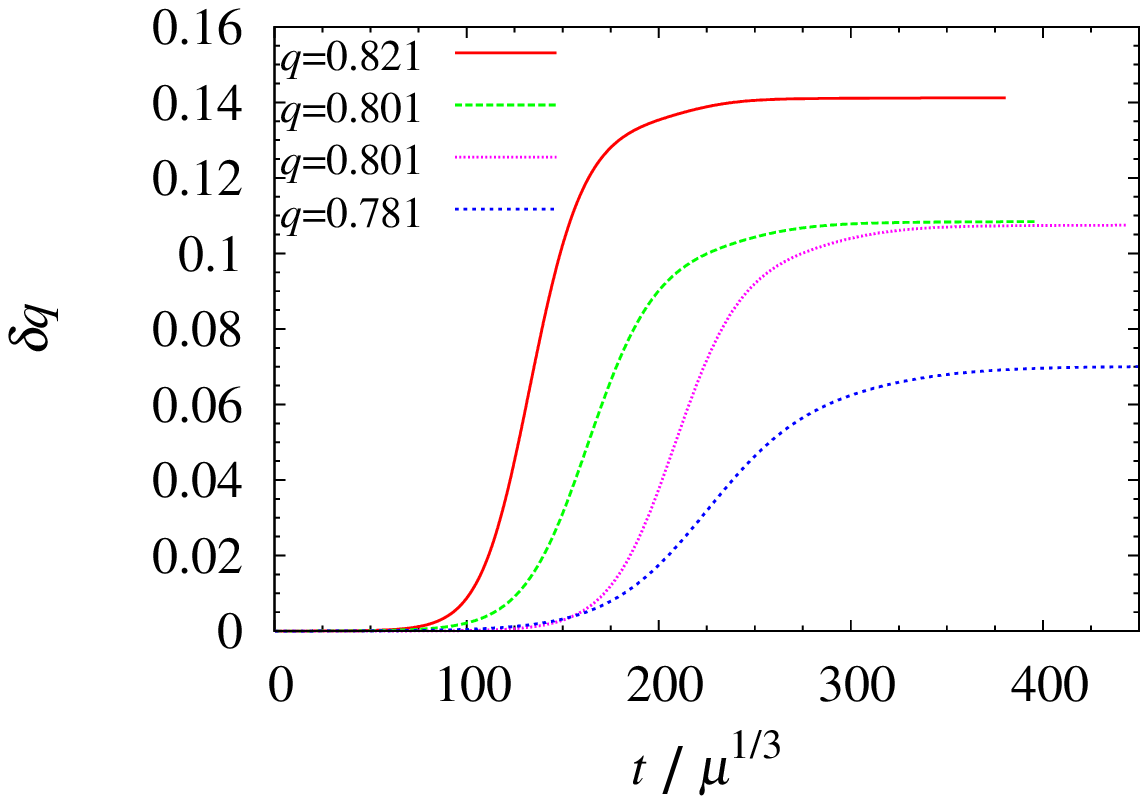}
\epsfxsize=3.3in
\leavevmode
~~~\epsffile{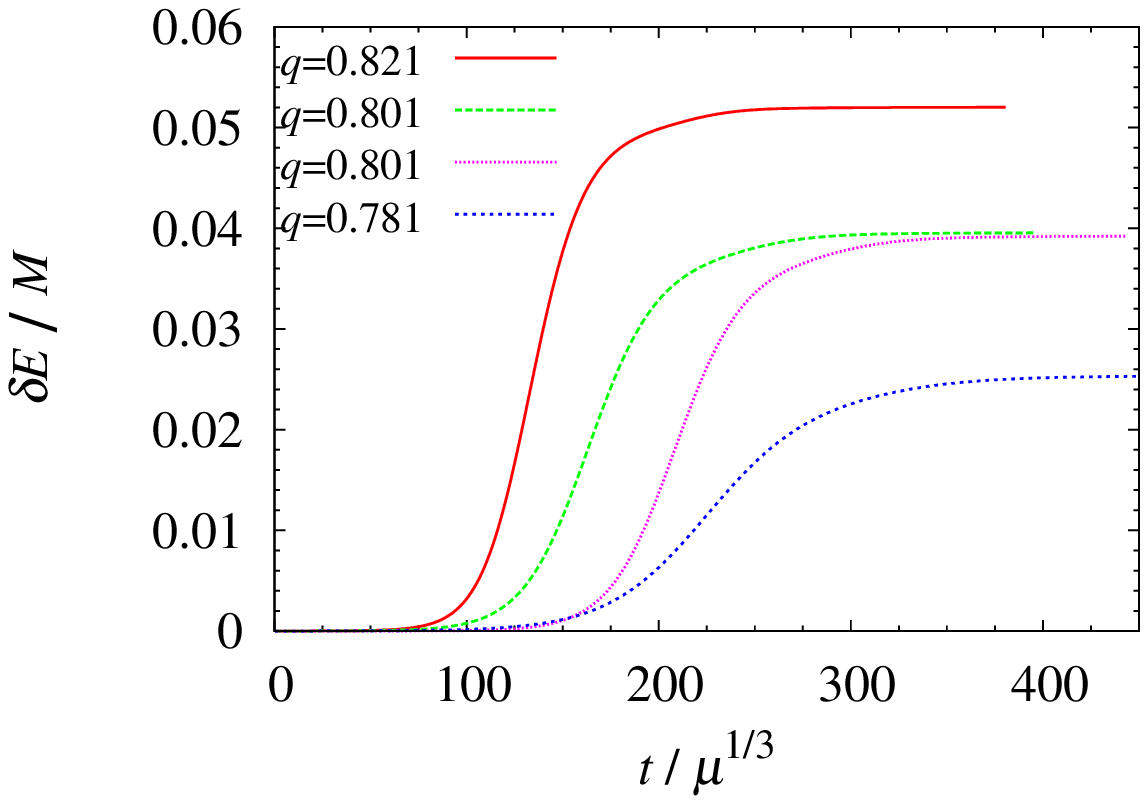}
\vspace{-5mm}
\caption{$|\delta q|$ (left) and $\delta E/M$ (right) as functions of $t$ 
for $d=6$. The results for $q_i=0.801$ with $A=0.005$ and 0.02, and
for $q_i=0.781$ and 0.821 with $A=0.02$ are displayed for $N=50$.  For
$q_i=0.821$, we stopped the simulation at $t/\mu^{1/3} \approx 370$
because the BH reaches approximately stationary state. 
\label{FIG9}} 
\end{figure*}

\subsection{Mechanism of spin-down}

Next, we consider the evolution of nondimensional spin parameter, $q$, 
as a result of gravitational-wave emission. 
The variation for this quantity, in the assumption of 
monochromatic gravitational-wave emission,  is written as
\beqn
\delta q &=&8 \pi G_d \mu^{-(d-2)/(d-3)} \Omega_{d-2}^{-1} 
\left( \delta J - {2a \over d-3} \delta E \right) 
\nonumber \\
&=&-8 \pi G_d \mu^{-(d-2)/(d-3)} \Omega_{d-2}^{-1} |\delta J| 
\left(1 - {2 \over d-3} {a\omega \over m} \right),~~~~
\label{deltaq1}
\eeqn
or
\beqn
\delta q=-{|\delta J| \over J}q 
\left(1 - {2 \over d-3} {a \omega \over m} \right).
\label{deltaq}
\eeqn
As found from Eq.~(\ref{cond0}), the gravitational-wave emission 
is possible only for $\Omega_{\rm H} > \omega/m$. Thus, 
\beqn
1 - {2 \over d-3} {a\omega \over m} &>& 1 - {2 \over d-3} a\Omega_{\rm H}
\nonumber \\
&=& 1 - {2 \over d-3} {a^2 \over r_+^2 + a^2}.
\eeqn
Because $2/(d-3) \leq 1$ and $a^2/(r_+^2 + a^2) < 1$ for any BH 
with $d \geq 5$, we find that $\delta q$ is always
negative for the superradiance mode.  Thus, the unstable BHs have to 
spin down by gravitational radiation reaction and evolve toward a 
stable state, as we found in numerical simulation (cf. Fig.~\ref{FIG6}).

We quantitatively confirm from Eq.~(\ref{deltaq}) that the spin-down of the
unstable BHs is indeed due to gravitational-wave emission as shown in
the following.  Using a formula of gravitational-wave luminosity
\cite{CDL03} (the Landau-Lifshitz pseudo tensor \cite{YS09} also gives 
the same formula after partial integration), the luminosity of
quadrupole gravitational waves ($l=|m|=2$ mode) is written as
\beqn
{dE \over dt}={(\dot h_+^2 + \dot h_{\times}^2) \Omega_{d-2} \over 16\pi G_d} 
{(d-3)d \over (d-2)(d+1)}\mu^{(d-2)/(d-3)}. \label{dedt0}
\eeqn
For monochromatic gravitational waves with angular 
frequency $\omega$, the luminosity may be rewritten as
\beqn
{dE \over dt}={\omega^2 (h_+^2 + h_{\times}^2) \Omega_{d-2} \over 16\pi G_d} 
{(d-3)d \over (d-2)(d+1)}\mu^{(d-2)/(d-3)}. \label{dedt}
\eeqn
In the following, we assume $h_+[\omega(t-r)]=h_{\times}[\omega(t-r)+\pi/2]$ 
(and set $h=\sqrt{h_+^2 + h_{\times}^2}$), 
and $\omega=$const for simplicity, which approximately hold as mentioned before.

Rewriting Eq.~(\ref{deltaq1}) in the form of spin evolution as
\beqn
{dq \over dt}=-{8\pi m G_d \mu^{-(d-2)/(d-3)} \over \omega \Omega_{d-2}}
\Big|{dE \over dt}\Big|
\left(1 - {2 \over d-3} {a \omega \over m} \right), \label{deltadqdt}
\eeqn
and integrating this equation, we may infer the spin-down history of a
BH for the whole evolution process by
\beqn
|\delta q(t)| = \int_0^t dt' {m\omega h^2 d(d-3) \over 2(d-2)(d+1)}
\Big(1 - {2 \over d-3}{a\omega \over m}\Big).\label{deltaqt} 
\eeqn
We also calculate dissipation history of the mass energy by
\beqn
\delta E(t) = \int_0^t dt' {dE \over dt},\label{deltaEt}
\eeqn
where we use Eq.~(\ref{dedt}) for $dE/dt$. We note that
Eq.~(\ref{deltaq1}) is valid only for the case that deformation from a
stationary BH solution is small. Thus, we here apply
Eq.~(\ref{deltaqt}) only for the small values of $q_i-q_{\rm crit} <
0.1$ for which the maximum value of $\eta$ is $\alt 0.15$ and the
deformation remains in a weakly nonlinear level.

Figure \ref{FIG9} plots $|\delta q|$ and $\delta E/M$ as functions of
$t$ for $d=6$ and for $q_i=0.821$, 0.801, and 0.781.  Here, we assume
that $a \omega =1.05q_i$ ($\omega \mu^{1/3}=1.05$) taking into account
the results shown in Fig.~\ref{FIG8}, and also neglect the dissipation 
of mass and spin of the BHs in computing the integrals of
Eqs.~(\ref{deltaqt}) and (\ref{deltaEt}), because the variation of
these quantities gives a minor effect on $\delta q$ and $\delta E$ for
$q_i-q_{\rm crit} < 0.1$; the error is likely to be less than 10\%.
As shown in Fig.~\ref{FIG9}, we obtain $\delta q_{\rm tot} \approx
0.14$, 0.11, and 0.07 for $q_i \approx 0.821$, 0.801, and 0.780,
respectively, which agree with the results determined from the final
value of $C_p/C_e$ within the error $\Delta(\delta q) \approx 0.006$.
Based on this good agreement, we conclude that the gravitational
radiation reaction determines the spin-down of the unstable BHs and
that in our simulations, the spin-down process is computed accurately.
We also note that the numerical results for the fractional increase of
the area (cf. Fig. \ref{FIG7}) agree well with that calculated from
$\delta q$ and $\delta E$ obtained here.

As we have described above, the evolution process of an unstable BH
with $q \alt 1$ is determined by gravitational radiation reaction.  As
far as the growth rate of the unstable mode is comparable with or
smaller than the spin-down rate due to gravitational radiation
reaction, the scenario presented here will be always correct. Figures
\ref{FIG1}(b) and \ref{FIG3}(b) show that for the unstable BHs with
$q_i \alt 1.1$, which have the growth timescale ($\tau$) longer than
$\pi/\Omega_{\rm H}$, this scenario holds.  We, however, note that
this scenario might not be valid for $q \gg 1$: If the growth
timescale is too short (e.g., $\tau \omega \alt 1$), the BH could not
emit a significant amount of gravitational waves in $\tau$ for
spinning down, and as a result, it can achieve a state for which the
deformation parameter is of order unity.  Extrapolation of the results
shown in Fig.~\ref{FIG2}(d) indeed suggests that for $q_i \agt 1.6$,
$\eta$ may reach $\agt 1$ for $d=6$. For such case, gravitational
radiation reaction may not prevent the growth of the bar-mode
deformation, and then, highly nonlinear deformation may lead to
fragmentation of the BH (in the classical argument, formation of a
naked singularity may be the result), as discussed in \cite{EM03}. In
the present work, we have not pursued the possibility that $\eta$
reaches $\sim 1$, because a longterm simulation for such an ultra
spinning BH is not an easy task technically. We leave such a work for
the future. In Sec.~\ref{sec:sum3}, we speculate a condition for an
ultra spinning BH to cause fragmentation.

Before closing this section, we note that the fraction of total
radiated energy of gravitational waves is much smaller than that of
angular momentum. The reason for this is that the following relation
holds:
\beqn
{\delta J \over J_0}={d-2 \over a\omega}{\delta E \over M_0}.
\eeqn
Here, $M_0$ and $J_0$ are initial mass and angular momentum of a BH.
For $d=6$ and $q_i=0.801$, for example, $\delta E/M_0$ is only
$\approx 4\%$, whereas $\delta J/J_0$ is $\approx 20\%$.  This
significant angular-momentum dissipation is essential for the
spin-down of the unstable BHs. It is worthy to note that the angular
momentum dissipation rate is even larger for the larger number of $d$
for a given value of $q$ because $\omega\mu^{1/(d-3)}$ depends weakly
on $d$ for the unstable BHs.


\section{Summary and discussions}

\subsection{Summary}

We show by numerical-relativity simulation that BHs spinning
sufficiently rapidly are unstable against nonaxisymmetric bar-mode
deformation for $d=6$, $7$ and $8$, as in the case $d=5$ \cite{SY09}.
In this instability, gravitational waves are emitted
spontaneously. The critical BH spin for the onset of the bar-mode
instability is $q_{\rm crit} \approx 0.74$ for $d=6$, $\approx 0.73$
for $d=7$, and $\approx 0.77$ for $d=8$, respectively. Thus, the
critical value is smaller than unity, and depends only weakly on the
dimensionality for $d \geq 6$.  After the instability sets in, the
degree of the bar-mode deformation increases exponentially with time,
and eventually, it saturates. The saturation is caused by
gravitational radiation reaction. For the larger initial spin, the
growth timescale is shorter. The degree of the maximum deformation is
larger for the larger initial value, approximately in proportional to
$q_i-q_{\rm crit}$. 

The unstable BHs emit gravitational waves significantly, and then,
spin down, settling to a stable BH with $q < q_{\rm crit}$.  The final
value of the BH spin, $q_f (< q_{\rm crit})$, is smaller for the
larger initial spin: For $d=6$ with $q_i-q_{\rm crit} < 0.1$, the
final spin is approximately written as $q_f = 2q_{\rm crit}-q_i$, and
for $0.1 \alt q_i-q_{\rm crit} \alt 0.3$, the final spin is in a
narrow range $q_f \approx 0.6$ -- 0.65. For $d=7$, $q_f \approx 0.66$ 
--0.69 for $0.85 \alt q_i \alt 1$. The smaller value of $q_f$ for
the larger initial spin $q_i$ is due to a larger amount of
gravitational-wave emission induced by the higher degree of the
bar-mode deformation. The results for $q_i \sim 1$ indicate that for
$q_i \gg 1$, the final spin is likely to be $\sim 0.6$ for $d=6$ and 
$\sim 0.65$ for $d=7$, if the BH settles to a stable state. 


The dynamically unstable BHs satisfy two conditions. One is that a
quasinormal mode satisfies a superradiance condition; the real part of
the eigen angular frequency for the quasinormal mode, $\omega =
\mathrm{Re}(\omega_{\rm QN})$, has to satisfy the condition $\omega <
m \Omega_{\rm H}$. The other is that the imaginary part of the eigen
angular frequency, $\omega_I = \mathrm{Im}(\omega_{\rm QN})$, is
negative.  The superradiance condition referred to here is satisfied
not only for the dynamically unstable BHs but also for the stable BHs
with the spin smaller than the critical value ($q_{\rm crit}$). The
minimum spin for which the real quasinormal frequency satisfies the
superradiance condition, $q_{\rm SR}$, is $\sim 0.6$ for $d=5$, $\sim
0.65$ for $d=6$, $\sim 0.7$ for $d=7$, and $\sim 0.75$ for $d=8$,
respectively. Thus, $q_{\rm SR}$ increases with the number of
$d$. This suggests that $q_{\rm crit}$ also increases with $d$ for $d
> 8$, because $q_{\rm crit}$ has to be larger than $q_{\rm SR}$ for
the spontaneous gravitational-wave emission.

It is worthy to note that the dynamically unstable BHs we found always
satisfy the above two conditions. If the superradiance condition were 
not satisfied, a BH could not emit gravitational waves spontaneously,
even if it is dynamically unstable ($\omega_I < 0$).  Probably, there
would exist a mathematical proof for the relation $q_{\rm crit} \geq
q_{\rm SR}$.

The value of $q_{\rm crit}$ depends weakly on the dimensionality for
$d \geq 6$ as mentioned above.  By contrast, this value is slightly
larger for $d=5$ ($q_{\rm crit} \approx 0.87$) \cite{SY09} and absent
for $d=4$. This shows that properties of the spinning BHs are
qualitatively similar for $d \geq 6$, whereas those for $d=4$ and 5
have their intrinsic properties.

\subsection{Axisymmetric vs nonaxisymmetric vs fragmentation
instabilities}
\label{sec:sum3}

The critical spin for the onset of the bar-mode instability found in
this paper is much smaller than that for the onset of axisymmetric
instabilities \cite{Dias,Murata}, $q \geq 1.56$ -- 1.80 for $d=6$ --
9, respectively (the minimum spins depend weakly on the
dimensionality).  Considering the analogy with the instabilities on
rotating stars, we find that this is a quite reasonable consequence
\cite{CH69,EH81,EH82,Cardoso}: Rapidly and rigidly rotating stars
can be unstable against axisymmetric instabilities (e.g., against
toroid or ring formation) only for an extreme case in which the axial
ratio of the polar axial length to the equatorial one is quite small
(e.g., smaller than 0.171 for the incompressible fluid \cite{EH81}).
By contrast, a variety of nonaxisymmetric instabilities can set in
even for the case that such ratio is not very small (for $\alt 0.58$,
secular instabilities set in, and for $\alt 0.31$, dynamical
instabilities set in for the incompressible fluid
\cite{CH69,EH82}). Also, the bar-mode instability occurs at the lowest
critical spin among many other instabilities for most rotating stars;
the bar-deformation is the most efficient way for decreasing total
energy of the system.  The bar-mode instability may be most relevant
for any self-gravitating spheroidal objects spinning rapidly.

According to an estimate based on the BH thermodynamics by Emparan and
Myers \cite{EM03}, the critical spin for the onset of a 
nonaxisymmetric instability is approximately unity, $q_{\rm frag}
\approx 1$, irrespective of the number of $d$ for $d \geq 6$. The
assumption in their argument is that a rapidly spinning BH of area
$A_i$ will fragment into two non-spinning boosted BHs of the total
area $A_f$ which is larger than $A_i$. For the instability found in this paper
for $q_i \alt 1$, the growth of the perturbation saturates at a weakly
nonlinear level at $\eta \alt 0.5$, because gravitational-wave emission
suppresses the further nonlinear growth. Thus, the fragmentation does
not occur for $q_i \alt 1$. However, this does not imply that the
fragmentation instability is not relevant for any spin. As we showed
in Figs.~\ref{FIG1} and \ref{FIG3}, the growth timescale of the
unstable mode depends strongly on the magnitude of the spin, and for
$q \agt 1$, it is shorter than a spin period of the BH, $\tau
< 2\pi/\Omega_{\rm H}$. For $q \gg 1$, the growth timescale is likely
to be even shorter. In such case, gravitational-wave emission will
play a minor role and nonlinear growth will continue until a state of
$\eta \sim 1$ is achieved, as discussed in Sec.~V.  Then, highly
nonlinear bar-mode deformation may lead to a fragmentation of the BH
due to the Gregory-Laflamme instability \cite{GL93}. In the following,
we infer how large initial spin is required for achieving the 
fragmentation focusing on the case $d=6$. 

As we showed in Figs.~\ref{FIG1} and \ref{FIG3}, the degree of bar-mode 
deformation for the unstable BHs increases exponentially with time as 
far as $\eta \alt 0.1$ as 
\beqn
\dot \eta ={\eta \over \tau},
\eeqn
where $\tau$ is a function of $q$, and for $d=6$, $\tau^{-1} \approx
C_{\tau}(q-q_{\rm crit})$ with $C_{\tau} \approx 0.51\mu^{-1/3}$. 
Gravitational radiation reaction plays a crucial role for
suppressing the growth of the deformation.  A nonlinearity associated
with this effect plays an important role for larger values of $\eta$.
We take into account this nonlinear effect by phenomenologically
replacing $\tau$ to $\tau (1 + C_s \eta /2)$ where $C_s$ is a constant
of order unity.  Because the amplitude of gravitational waves, $h$, is
approximately written as $\eta/2$, we may assume the following
approximate equation for the growth of gravitational-wave amplitude:
\beqn
\dot h ={h \over \tau (1 + C_s h)}.
\eeqn
Combining this relation with Eq.~(\ref{deltadqdt}), we obtain
\beqn
&&(1 + C_s h){d h^2 \over d q} \nonumber \\
&& ~~~~~=-{4 \over \tau m \omega}{(d-2)(d+1) \over (d-3)d}
\Big(1-{2 \over d-3}{a \omega \over m}\Big)^{-1}. 
\eeqn
This equation is valid only for $h \ll 1$, but in the following, we
use it even for $h=O(1)$ to infer the maximum value of $h$ as a
function of $q_i$.  For $d=6$ with $m=2$ and $q_{\rm crit} < q \alt
1.1$, this relation is approximately written as
\beqn
(1 + C_s h){d h^2 \over d q} \approx -C_{\tau}(q-q_{\rm crit}){28 \over 9 \omega}
\Big(1-{a \omega \over 3}\Big)^{-1}. 
\eeqn
Because the factor, $1-a\omega/3$, is a slowly varying function of $q$
in the range 0.76 -- 0.64 for $q=0.7$ -- 1.1, we assume that it is a
constant and denote it by $F^{-1}$. In addition, we assume $\omega
\mu^{1/3} \approx 1$ for simplicity. Using the approximate relation
for $C_{\tau}$, we finally obtain
\beqn
(1 + C_s h){d h^2 \over d q} \approx -1.6 F(q-q_{\rm crit}). \label{hq}
\eeqn 
Here, we assume that the maximum value of $h$ ($h_{\rm peak}$) is
achieved when the spin $q$ reaches $q_{\rm crit}$ as a result of
spin-down. Then, integrating Eq.~(\ref{hq}) from $q=q_i$ to $q=q_{\rm
  crit}$ gives
\beqn
h_{\rm peak}^2 + {2C_s \over 3}h_{\rm peak}^3
 \approx 0.8F(q_i-q_{\rm crit})^2, \label{etam}
\eeqn
or, using $h_{\rm peak} \approx \eta_{\rm max}/2$, 
\beqn
\eta_{\rm max}^2 + {C_s \over 3}\eta_{\rm max}^3
 \approx 3.2F(q_i-q_{\rm crit})^2. \label{etam2}
\eeqn
Here, $3.2F$ is $\approx 4.2$ -- 5.0 for $q \approx 0.7$ -- 1.1. 

Equation (\ref{etam2}) is indeed a good approximate relation for an 
appropriate choice of $C_s$.  The solid curve of Fig.~\ref{FIG2} (d) is 
\beqn
\eta_{\rm max}^2 + 2\eta_{\rm max}^3=4(q_i-q_{\rm crit})^2, 
\label{etamaxf}
\eeqn
and fits the numerical results well. It is found that for a high spin
$q_i \agt 0.9$, the nonlinearity associated with gravitational
radiation reaction [the second term in the left-hand side of
  Eq.~(\ref{etamaxf})] plays an important role in determining the
value of $\eta_{\rm max}$.

Assuming that Eq.~(\ref{etamaxf}) is valid even for $q_i > 1.1$, we
expect the maximum value of $\eta$ and find that $\eta_{\rm max}
\agt 1$ for $q_i \agt 1.6$ (note that the possible maximum value of
$\eta$ is 2).  This suggests that the fragmentation may occur for an
ultra spinning BH and the critical spin for the fragmentation, $q_{\rm
  frag}$, is $\agt 1.6$; this predicted critical value is much larger
than the value predicted in \cite{EM03}. The critical spin for the
onset of axisymmetric instabilities is fairly close to $q_{\rm frag}$
as $q_{\rm axis} \approx 1.56$ for $d=6$ \cite{Murata}.  Therefore,
for the ultra spinning case, the BH may fragment in a complicated
manner.  We note that for $\eta_{\rm max} \agt 1$, several additional
nonlinear effects will play an important role for the evolution of the
deformed BH. For example, emission of gravitational waves with modes
other than the lowest-order quadrupole mode could contribute to the
spin-down and to suppressing the growth of the bar-mode instability.
Thus, the critical spin for the fragmentation may be even larger than
1.6. Finally, we note that essentially the same argument holds also
for $d \geq 7$. 

\subsection{Implications for mini black hole evolution}

The bar-mode instability found in this paper changes the hypothetical
picture for the evolution of a mini BH which may be formed in particle
accelerators. It is natural to expect that most of the mini BH is
formed for a large impact parameter, which is close to a critical
value, $b_{\rm max}$, in two-particle collision, and that the BH
should be rapidly spinning at its formation.  In fact,
analyses~\cite{YN03,YR05} for high-velocity two-particle collisions
indicate that a BH can be formed for impact parameters such that the
resulting BH is rapidly spinning with $q > 1$ for $d \geq 6$.  To
clarify this point, we first review the results of an analysis for
high-velocity particle collision with the impact parameter $b$ and
with the energy of each incoming particle $p$~\cite{YN03,YR05}.  For
this phenomenon, the total mass energy and angular momentum of the
system is $M = 2p$ and $J=bp$, respectively, assuming that the
particles move with the speed of light.  As in the case of the
Myers-Perry BH of single spin, we define spin and mass parameters, $a$
and $\mu$, for this system using the same formulas as
Eqs.~\eqref{MP-parameter-mu} and \eqref{MP-parameter-a}, and then
calculate nondimensional spin $q=a/\mu^{1/(d-3)}$.  Table~\ref{table1}
shows the maximal impact parameters $b_{\rm max}/\mu^{1/(d-3)}$ for
the apparent horizon formation obtained in \cite{YR05} and the
corresponding value of the spin $q=q_{\rm max}$ for $d=5$ -- 8. The
value of $q_{\rm max}$ is larger than unity for $d=6$ -- 8, and thus,
the resulting BH is likely to be rapidly spinning so that it can be
subject to the dynamical bar-mode instability
\footnote{In this analysis, we assume that the mass and angular
  momentum for the formed BH are equal to those of the system for
  simplicity.  However, this is not likely to be the case because
  gravitational waves should be significantly emitted during the
  collision. The value of $q_{\rm max}$ is nothing but a roughly
  approximated value. As we showed in this paper, angular momentum
  will be more efficiently emitted by gravitational waves than
  energy. Thus, the resulting spin could be smaller than $q_{\rm max}$
  in reality.}.

\begin{table}[tb]
\caption{The values of maximal impact parameter $b_{\rm max}/\mu^{1/(d-3)}$
for the apparent horizon formation obtained in \cite{YR05} 
and the corresponding nondimensional spin of the system, $q_{\rm max}$. 
}
\begin{ruledtabular}
\begin{tabular}{c|cccc}
 $d$ &  $5$ &$6$ & $7$ & $8$  \\
  \hline
 $b_{\rm max}/\mu^{1/(d-3)}$ & $1.24$ &$1.47$ & $1.59$ & $1.66$ \\
 $q_{\rm max}$ & $0.93$ &$1.47$ & $1.98$ & $2.50$ \\
  \end{tabular}
\end{ruledtabular}
\label{table1}
\end{table}

The phenomenology of a mini BH formed in particle accelerators is
determined by two timescales: One is the timescale of 
gravitational-wave emission $\tau_{\rm GW} \sim 
1/|\omega_I|$. The other is the timescale of Hawking radiation
$\tau_{\rm H}$.  For convenience, we denote these timescales in terms
of the Planck mass $m_{\rm P}$. Among several manners of defining
the Planck mass (summarized in \cite{GT02}), we adopt the
definition as 
\begin{equation}
m_{\rm P} = \left[\frac{(2\pi)^{d-4}}{4\pi G_d}\right]^{1/(d-2)}.
\end{equation}
Then, the Planck time is defined by $\tau_{\rm P} := 1/m_{\rm P}$ 
in the natural units.

The timescale of gravitational-wave emission is written as
\begin{equation}
\tau_{\rm GW} 
= C_{\rm GW} \tau_{\rm P}
\left({M_0 \over m_{\rm P}}\right)^{1/(d-3)},
\end{equation}
where $M_0$ is the initial BH mass.  If the formed BH is stable, 
$\tau_{\rm GW}$ should be defined by $1/|\omega_I|$, and the
coefficient is
\begin{equation}
C_{\rm GW} = \frac{2\pi}{\omega_I}
\left[\frac{2}{\pi(d-2)\Omega_{d-2}\mu}\right]^{1/(d-3)}. \label{cgw}
\end{equation} 
For the 6D head-on collision ($q=0$), for example, $C_{\rm GW}\approx
3.8$ \cite{YSS}, and this factor depends weakly on the dimensionality
as long as $q=0$.  For spinning BHs, it becomes larger, and $C_{\rm GW}
\gg 1$ for $q \sim q_{\rm crit}$ as found in the present paper; 
practically $C_{\rm GW}=O(100)$, 
if gravitational radiation reaction is taken into account.

If the value of $q$ is further increased to be $q > q_{\rm crit}$, 
the value of $C_{\rm GW}$ should be approximately written as 
\begin{equation}
C_{\rm GW} = \frac{2\pi}{|\omega_I|}
\left[\frac{2}{\pi(d-2)\Omega_{d-2}\mu}\right]^{1/(d-3)} + C_{\rm GW0},
\end{equation} 
where the first term in the right-hand side is associated with the
growth timescale for the bar-mode deformation and the second one is
the timescale of damping after the saturation is achieved, which is
approximately equal to $C_{\rm GW}$ in Eq.~(\ref{cgw}) for $q \alt
q_{\rm crit}$ and thus of order 10 -- $10^2$ [cf. Fig.~\ref{FIG2}(b)
  and (c)].  The first term is comparable to the second term for
$q_{\rm crit} < q \alt q_{\rm crit} + 0.1$ but the second term
dominates for the larger value of $q$.

The timescale of Hawking radiation is
\beqn 
\tau_{\rm H} = C_{\rm H} \tau_{\rm P}
\left({M_0 \over m_{\rm P}}\right)^{(d-1)/(d-3)}, 
\eeqn 
where $C_{\rm H}$ is a $q$-dependent constant and determined taking
into account all emission processes (i.e., emission of scalar, spinor,
vector particles and gravitons).  For the Schwarzschild case, $C_{\rm
  H}$ can be evaluated using numerical results of the greybody factor
\cite{greybody}; e.g., $C_{\rm H}\approx 1.7$ for $d=6$.  The greybody
factors for spinning BHs are studied for brane spacetimes 
\cite{Ida1,Ida2,Ida3,greybody-rotating} whereas those for bulk
gravitons have never been completely evaluated (but see
\cite{KKKPZ09}).  The general tendency is that the emission rate is
enhanced significantly as the value of $q$ or $d$ is increased: The
luminosity for $q \sim 1$ is $\sim 100$ times as large as that for
$q=0$ (e.g., \cite{Ida3}).  The angular momentum emission rate is more
significantly enhanced as the spin is increased, indicating that a 
rapidly spinning BH spins down in a short timescale.  Thus,
strictly speaking, we should take into account two timescales: One is
the evaporation timescale $\tau_{\rm H}$ and the other is the
spin-down timescale referred to as $\tau_{\rm Hs}$.  Reference
\cite{Ida3} shows that for $q \sim 1$, $\tau_{\rm Hs} \ll \tau_{\rm
H}$: $\tau_{\rm Hs}$ for $q=1$ is by a factor of $\sim 100$ shorter
than that for $q \ll 1$, but $\tau_{\rm H}$ is not as short as $\tau_{\rm
Hs}$.

The ratio of $\tau_{\rm H}$ to $\tau_{\rm GW}$ is  
\beqn 
{\tau_{\rm H} \over \tau_{\rm GW}} = {C_{\rm H} \over
  C_{\rm GW}} \left({M_0 \over m_{\rm P}}\right)^{(d-2)/(d-3)}, 
\eeqn 
and proportional to $M_0^{(d-2)/(d-3)}$ where $1 < (d-2)/(d-3) \leq
4/3$ for $d \geq 6$. Thus, the ratio depends moderately on $M_0$ for
$m_{\rm P} < M_0 \alt 10 m_{\rm P}$. On the other hand, $C_{\rm
H}/C_{\rm GW}$ are in the range $\sim 10^{-3}$ -- 1, depending
strongly on the spin.  This implies that the evolution of a mini BH formed
after a particle collision depends strongly on the spin $q$ (or
equivalently, the impact parameter $b$).

Now, we discuss the phenomenology of a mini BH assuming that $m_{\rm
  P} = 1~\mathrm{TeV}$ and $M_0=10~\mathrm{TeV} =10 m_{\rm P}$, which
are plausible values in TeV-gravity scenarios and in LHC. Note that the 
scenario is qualitatively unchanged as far as $M_0 \alt 10^2 m_{\rm
  P}$. As a specific example, we further fix $d=6$ in the following. 
For $M_0 = 10 m_{\rm P}$, $\tau_{\rm GW}\approx
8 \tau_{\rm P}$ and $\tau_{\rm H}\approx 80 \tau_{\rm P} \gg \tau_{\rm
  GW}$ for $q=0$. As the value of $q$ is increased from
$q=0$, $\tau_{\rm GW}$ becomes longer and $\tau_{\rm H}$ shorter (for
a small value of $q \ll 1$, $\tau_{\rm Hs}$ is comparable to
$\tau_{\rm H}$).  Thus, two timescales, $\tau_{\rm GW}$ and $\tau_{\rm
  H}$, become identical, $\tau_{\rm GW}=\tau_{\rm H}$, at a value of
$q=q_{\rm eq}$.  The result in this paper indicates $q_{\rm eq}<
q_{\rm crit}$, because $\tau_{\rm GW}$ becomes very long ($C_{\rm GW}$
is of order 100) for $q \sim q_{\rm crit}$.

Then, the phenomenology is classified into three types depending on
the spin $q$; (i) $0 \leq q \lesssim q_{\rm eq}$, (ii) $q_{\rm eq
}\lesssim q < q_{\rm frag}$, and (iii) $q \geq q_{\rm frag}$.  Here,
we assume a hypothetical critical value for the onset of fragmentation
as $q_{\rm frag} \agt 1.6$ following the estimate in Sec.~VI~B,
although it is not clear whether the fragmentation really
occurs. Because of the high value of $q$, the fragmentation will
proceed in a timescale of order $M_0^{1/(d-3)} \sim \tau_{\rm H} \ll
\tau_{\rm GW}$, if it occurs.

For the case (i), the standard hypothetical picture of BH evaporation
holds; a formed BH emits gravitational waves and settles to a
stationary state in a short timescale. Then, it will be evaporated by
Hawking radiation.  The formed BH is spinning, but not very rapidly,
and the effect of the energy loss and spin-down by gravitational
radiation reaction plays a minor role. The spin-down will proceed
primarily by Hawking radiation, and finally, a nearly nonspinning BH
will be evaporated. This type of BH formation and evaporation will not
be a dominant process in particle accelerators, because a mini BH with
such a small spin ($q < q_{\rm eq} < q_{\rm crit}$) or with a small
impact parameter will not be formed frequently; the formation rate
will be by a factor of $(q_{\rm eq}/q_{\rm max})^2$ smaller than the total. 

For the case (ii), the BH will not be relaxed to a stable state by
gravitational-wave emission, because of its long emission timescale.
Such a nonstationary BH will start radiating quantum particles while 
emitting gravitational waves, and the signal of the quantum radiation
in this phase is likely to be different from the idealized Hawking
radiation (which is the result only from a stationary, axisymmetric
BH). Because a BH with a large initial spin reaches a highly deformed
state, non-Hawking-type quantum radiation will be enhanced for the
higher spin. Obviously, an improved analysis for predicting the
spectrum of the quantum radiation is required. Assuming that the
spin-down timescale by the quantum radiation is shorter than the
evaporation timescale as in the Hawking radiation \cite{Ida3}, the BH
spin will subsequently become smaller than $q_{\rm crit}$ after the
substantial quantum radiation, and thus, the BH relaxes eventually to
a stable, quasistationary, axisymmetric state and will stop emitting
gravitational waves.  After the condition $q < q_{\rm crit}$ is
achieved, the quantum radiation process will be identical to the
ordinary Hawking radiation.

For the case (iii), the timescale for the growth of bar-mode
deformation and subsequent fragmentation is likely to be of order
$\sim M_0^{1/(d-3)}$. This may be as short as $\tau_{\rm Hs}$. If
fragmentation occurs, a rapidly spinning BH is likely to change to
two slowly spinning BHs before the quantum (Hawking) radiation becomes
the dominant dissipation process.  A supportive evidence for this
possibility is that the apparent horizon formed in particle collisions
takes a peanut shape for $b\approx b_{\rm max}$ as shown in Figs.~5
and 6 of \cite{YR05}, which indicates that gravity combining the two
particles is weak.  If the fragmentation occurs (the BH horizon
pinches off) because of quantum-gravity effect, two boosted BHs will
be the outcome (binary will not be the result for $d \geq 5$). Thus,
the Hawking radiation from the two boosted BHs may be observed
subsequently, as two jets in particle accelerators.  Here, we note
that the two timescales, $M_0^{1/(d-3)}$ and $\tau_{\rm Hs}$, may be
as short as the Planck time $\tau_{\rm P}$ for $M_0 \approx 10 m_{\rm
  P}$, and thus, such a (semi) classical phenomena may be veiled by
quantum-gravity effects in LHC. 

\subsection{Issues for the future}

The present paper reports the results for the stability of rapidly
spinning BHs against bar-mode deformation. We analyzed the BHs only
with the spin $q \alt 1.15$ and did not study for the ultra spinning
case with $q \agt 1.5$ because it is not technically easy to perform a
longterm simulation for such BHs. As discussed above, the evolution of
the ultra spinning BHs may be qualitatively different from that for $q
\alt 1$. Furthermore, such ultra spinning BHs may be the frequent
outcomes, if the TeV-gravity hypothesis is correct. Clarifying the
evolution of the ultra spinning BHs is obviously an important issue
left for the future.

The stability of black rings against bar-mode deformation is one of
the interesting issues. The analytic solutions for the black rings are
found in five dimensions \cite{ER03}, and numerical-relativity
simulation in a similar manner to that in this paper may be
possible. The black rings always have a high spin with $q > q_{\rm
crit}$ for $d=5$. Perhaps, they are also unstable against bar-mode
deformation and evolve as a result of gravitational radiation reaction
as far as $q$ is not extremely large; if the spin is very large, 
fragmentation may occur. 

In the present paper, we start simulations in a highly idealized
situation; we prepare nearly stationary, axisymmetric BHs and evolve
them approximately in a quasistationary manner. In particle
accelerators, however, the situation will be highly different. The
rapidly spinning BHs after the particle collision will be highly
nonstationary and nonaxisymmetric. Such BHs may evolve qualitatively
in a similar manner to that in the present analysis, but quantitative
properties on the evolution process will be significantly
different. To clarify the formation and evolution of such BHs, it is
necessary to perform a simulation started from high-velocity two-BH
collision. Our goal is to successfully perform this simulation and to
clarify the possible outcome in this setting. 

\acknowledgments 

We thank T. Tanaka, T. Shiromizu, M. Sasaki, and H. Kodama for
discussions. HY thanks also V.P.~Frolov and A.~Zelnikov for comments.
Numerical computations were in part performed on the NEC-SX9 at CfCA
in National Astronomical Observatory of Japan and on the NEC-SX8 at
Yukawa Institute for Theoretical Physics in Kyoto University. This
work was in part supported by Grant-in-Aid for Scientific Research
(21340051) and by Grant-in-Aid for Scientific Research on Innovative
Area (20105004) of the Japanese MEXT.  HY is supported by JSPS.

\end{document}